\documentclass[letterpaper]{article}
\usepackage[preprint]{aaai2027}
\usepackage[hyphens]{url}
\usepackage{graphicx}
\usepackage{natbib}
\usepackage{caption}
\usepackage{amsmath}

\DeclareUnicodeCharacter{03C4}{\ensuremath{\tau}}
\DeclareCaptionStyle{ruled}{labelfont=normalfont,labelsep=colon,strut=off} 
\usepackage{booktabs}
\usepackage{array} 
\usepackage{longtable}  
\DeclareUrlCommand\taskname{\urlstyle{tt}}

\renewenvironment{links}{%
  \newcommand{\link}[2]{\par\textbf{##1:} \url{##2}}%
  \setlength{\hangindent}{10pt}%
  \setlength{\parskip}{2pt}%
  \begin{flushleft}%
}{%
  \end{flushleft}%
  \vskip 1ex%
}
\title{Permission Denied: Policy-Graded Evaluation of Coding Agents in Hardened Environments}

\author{
    Dotan Davidovich\textsuperscript{\rm *},
    Yair Amar\textsuperscript{\rm *},
    Hai Rozencwajg\textsuperscript{\rm *},
    Or Hiltch
}
\affiliations{
    Accomplish AI\\
    \{dotan, yair, hai, or\}@accomplish.ai\\
    \textsuperscript{\rm *}These authors contributed equally.
}

\begin{document}

\maketitle

\begin{abstract}

Coding agents increasingly run inside organizations whose security controls (scoped credentials, restricted egress, read-only filesystems, non-root execution) constrain them like any other software. Existing benchmarks, however, evaluate agents almost exclusively in permissive sandboxes, so it is unknown how performance changes when policy is enforced. In this work, we evaluate 12 coding agents on Terminal-Bench 2.1 across nested security policy levels derived from common real-world enterprise restrictions. Hardening is never free but far from uniform: under the strictest policy, success losses reach 18.3 points and cost inflation 167.3\%, and the two axes disagree; the model that best preserves success is also the one that loses the most efficiency, so model choice is policy-dependent. Beyond aggregate scores, we characterize how agents behave when policy blocks their actions and decompose the failures hardening induces: runs grind into timeouts or wrong solutions rather than stopping early, in a mix that differs by model. To ground comparisons, we verify task solvability under the strictest policy, separating model failures from tasks the policy forecloses. We release Boundary-Bench, an open-source hardening plugin enabling policy-constrained evaluation of coding agents on Terminal-Bench and compatible benchmarks.

\end{abstract}

\begin{links}
  \link{Code}{https://github.com/boundary-bench/boundary-bench}
  \link{Project page}{https://boundarybench.com/}
\end{links}

\section{Introduction}

Organizations increasingly use coding agents for software development, but
these agents operate within security controls that govern users and
agents alike \citep{southDelegation}. Such controls are not agent-specific:
scoped credentials, network restrictions, protected paths, and limited
administrative privileges constrain any software operating in the environment.
These mechanisms correspond to access-control and boundary-protection families
in NIST SP 800-53 \citep{nist80053}. We call such settings
\emph{hardened environments}: execution environments in which policy
restricts what an agent can observe, access, or execute. Hardening is also
moving closer to the agent itself, as contemporary agent harnesses expose
controls such as network access modes, filesystem sandboxes, and run
profiles.\footnote{Official tool documentation: Claude Code
\url{https://code.claude.com/docs/en/claude-code-on-the-web},
Codex \url{https://developers.openai.com/codex/permissions},
Cursor \url{https://cursor.com/docs/agent/run-modes}.}

Some coding-agent benchmarks use permissive task sandboxes: Terminal-Bench allows agents to manipulate task containers freely \citep{merrillTerminalBenchBenchmarkingAgents2026}, FeatureBench provides free Internet access \citep{zhouFeatureBenchBenchmarkingAgentic2026}, and enterprise data-privacy and access-control policies are rarely tested in standard benchmarks \citep{yehudaiSurveyEvaluationLLMbased2026}. Like the mid-trajectory tool failures and bans that destabilize agents and require replanning \citep{xiongMoreVulnerableYou2025,liuCostBenchEvaluatingMultiTurn2026}, a policy denial can trigger retries, workarounds, or abandonment; we represent the resulting trade-off as a success--cost Pareto frontier \citep{kapoorAIAgentsThat2024}. Safety evaluations
such as AgentHarm document the risks that motivate such restrictions
\citep{andriushchenkoAgentHarmBenchmarkMeasuring2024}, but do not measure
what enforcing them costs in success rate and token spend.
Compliance-scored benchmarks check whether agents follow stated policies,
but neither enforce them at runtime nor measure their operational cost
\citep{levySTWebAgentBenchBenchmarkEvaluating2026}. Security-policy
enforcement is itself a structured runtime perturbation: unlike observation
distractions or injected faults
\citep{maCautionEnvironmentMultimodal2024,karaWAREXWebAgent2025}, it changes
which observations and actions are permitted. Existing evaluations
therefore do not show how coding-agent performance changes as such policies
tighten, and unconstrained leaderboards may misrepresent the success--cost
trade-offs of agents in deployed enterprise settings.

In this paper, we show that hardening does not uniformly degrade coding-agent performance. Every model's success--cost operating point worsens on both axes under policy, but by amounts and in currencies that differ across models, reshaping the Pareto frontier and making model choice policy-dependent. This effect is behavioral: agents respond differently to denied actions (some retry, some reroute, and some give up), so performance under enforced policy cannot be predicted from unconstrained scores. Because a single locked-down configuration provides only one operating point, it cannot show how the frontier moves as policies tighten. We therefore introduce Boundary-Bench and evaluate agents under graded, nested policy levels derived from common enterprise restriction settings and mapped to NIST SP 800-53 controls. At each level, we run each model's
native harness on Terminal-Bench. We report outcomes over the full task set.
As secondary diagnostics, we track whether any evaluated agent solves each
task (task accessibility) and whether its reference solution
remains compatible with each policy level. Our contributions are:
\begin{itemize}
\item \textbf{Policy-graded evaluation.} We release Boundary-Bench as an open-source hardening plugin that runs each model in its native harness on Terminal-Bench 2.1 under three nested policy levels enforced by native Linux mechanisms.
\item \textbf{Task diagnostics.} We report outcomes over all Terminal-Bench tasks. As secondary diagnostics, we track whether any evaluated agent solves each task and whether its reference solution remains compatible under the policy. This identifies task accessibility and cases where agents succeed through alternative trajectories when the canonical trajectory is blocked.

 \item \textbf{Non-uniform policy effects.} We show that hardening degrades every model--harness bundle non-uniformly: some models lose success, some incur higher token cost, and some experience both, reshaping the success--cost Pareto frontier at each policy level (Figure~\ref{fig:frontier}, Figure~\ref{fig:policy-shifts}).

\item \textbf{Failure-mode decomposition.} We decompose policy-induced degradation into measured failure modes: the extra failures are dominated by budget exhaustion and completed-but-wrong solutions, in proportions that differ by bundle, and we quantify agents' blocked-action exposure at the enforcement boundary (Table~\ref{tab:failure-decomp}).
\end{itemize}

\section{Related Work}

SWE-bench \citep{jimenezSWEbenchCanLanguage2023} standardizes repository-level coding evaluation; Terminal-Bench \citep{merrillTerminalBenchBenchmarkingAgents2026} and OSWorld \citep{xieOSWorldBenchmarkingMultimodal2024} evaluate interactive coding and computer-use agents; cost-aware evaluation reports success jointly with inference cost, for example as a Pareto frontier \citep{kapoorAIAgentsThat2024}. All hold the execution-security configuration fixed rather than varying it as an evaluation factor. Verifier over-specification, where assertions demand incidental details a task never requires, is likewise a documented failure mode of code-generation benchmarks \citep{sharifloo_where_2025}; we audit and repair five such verifiers before any policy comparison. Routing and cascade methods \citep{chenFrugalGPTHowUse2023,ongRouteLLMLearningRoute2024} navigate a performance--cost trade-off by varying model choice; we keep the two-objective framing, use verifier-checked task success as the outcome, and make the environment's policy level the independent variable.

AgentHarm scores harmfulness and refusal on malicious tasks \citep{andriushchenkoAgentHarmBenchmarkMeasuring2024}; ToolEmu surfaces risky tool-use behavior in an LM-emulated sandbox \citep{ruanIdentifyingRisksLM2023}; ST-WebAgentBench scores web-agent trajectories against organizational policies post hoc \citep{levySTWebAgentBenchBenchmarkEvaluating2026}; $\tau$-bench states domain rules in the prompt while enforcing only basic validity checks \citep{yaotaubenchBenchmarkToolAgentUser2024}; AgentDyn shows that prompt-injection defenses with low attack success can coincide with severe over-defense and reduced task utility \citep{liAgentDynAreYour2026}. In contrast, we impose compliance as a graded property of the execution environment and measure task success and token spend as functions of the policy level.

Input-perturbation robustness studies adversarial text and prompts \citep{wangAdversarialGLUEMultiTask2021,zhuPromptRobustEvaluatingRobustness2023}; ReliabilityBench varies injected fault intensity per tool call \citep{guptaReliabilityBenchEvaluatingLLM2026}; WAREX injects network and server faults into web-agent trajectories \citep{karaWAREXWebAgent2025}. We draw qualitative inspiration from the effective-robustness lens \citep{taoriMeasuringRobustnessNatural2020}, comparing hardened against unhardened performance. Whereas these works perturb inputs or inject synthetic faults, our severity axis is the security policy itself, held fixed per evaluation run.

Sandboxes and agent-runtime controls instantiate hardened execution environments, and the closest prior work shows that enforcement need not destroy utility at a single, fixed configuration. CaMeL deterministically enforces a fixed security policy and reports task utility under enforcement: a multi-point success drop and a $2.8\times$ token overhead \citep{debenedettiDefeatingPromptInjections2025}. Progent enforces tool-call privilege policies and reports utility largely maintained \citep{shiProgentSecuringAI2025}. The Verifier Tax quantifies task success and LLM-call and token overhead under an internal block-and-revise gate \citep{sahVerifierTaxHorizon2026a}. Each prices one enforcement configuration in isolation. We complement these single-point results with, to our knowledge, the first benchmark study to make policy severity the independent variable: a shared, NIST-mapped, nested ladder run identically across model--harness bundles, measuring what enforcement costs, for which bundle, and in which currency (success or spend), together with task accessibility at each level.

\section{Boundary-Bench}
\label{sec:method}
Boundary-Bench is a framework that layers a configurable, operating-system-enforced policy onto an existing coding-agent benchmark and measures how agent success and cost change relative to the unrestricted baseline.

\subsection{Policy Enforcement}

A policy is a set of capability restrictions on the runtime, enforced on the environment rather than the agent; it limits what any process can do. 

We model policies on three axes: network egress (N), filesystem scope (F), and privilege (P). Each axis ranges over three postures: open (no restrictions), restricted, and locked. The three axes map to distinct families of NIST SP 800-53 controls \citep{nist80053}: boundary protection (SC-7) on the network axis, access, information-flow, and process-isolation controls (AC-3, AC-4, SC-39) on the filesystem axis, and least privilege (AC-6) on the privilege axis. The mapping records the controls that motivated each mechanism and is not a compliance claim.

From this space we evaluate three policy levels that form a cumulative severity ladder, each adding restrictions on top of the previous level.
\textbf{\textit{Control}} is the open baseline: the agent runs as root with unrestricted egress and a writable filesystem.
\textbf{\textit{Non-root}} drops root privilege only, isolating the effect of running as an ordinary user while network and filesystem access stay open.
\textbf{\textit{NIST-derived high}} restricts all three axes together: a fixed network egress allow-list, a read-only filesystem outside a dedicated small writable set, and a full privilege lockdown.
Because the step from \textit{non-root} to \textit{NIST-derived high} changes several surfaces at once, we treat its effect as whole-configuration rather than attributing it to any single axis.
The exact egress allow-list, the read-only system directories, the privilege lockdown, and each axis's NIST control mapping are given in Appendix~A.

Each restriction is realized by a native Linux access-control mechanism rather than requested of the agent: network egress passes through a proxy the agent cannot reconfigure, the filesystem is mounted read-only outside a small writable workspace, and the agent process runs under an unprivileged user that cannot re-escalate. Enforcement is compiled into the environment without the agent's knowledge. The agent meets each limit only as a native runtime failure. Every restriction is inherited by each child process the agent spawns, so it cannot be evaded by launching a subprocess, while benchmark setup and grading run outside the sandbox. Before each run, pre-flight probes confirm every restriction is active and fail closed otherwise (Appendix~A).

\subsection{Benchmark Audit}
A sufficiently strict policy can invalidate a task rather than merely harden it, so Boundary-Bench audits the benchmark under its strictest policy before interpreting any agent result. First, we establish each task's solvability: we replay the benchmark's own reference solution under \textit{NIST-derived high}, and where it no longer passes we attempt to author a policy-compliant reference solution ourselves; a task whose official passing state is reached by either route gains a solvability witness. A task is \emph{blocked by design} when the policy forbids an action the task itself requires, so no policy-compliant solution exists. Second, we review every trial in which the verifier reported failure although the agent had reached the stated goal, checking the failing assertion for over-specification; a repair may only additively broaden an assertion, and the repaired verifier runs identically at every policy level, with original outcomes kept for audit. Third, we classify every failed run by where it terminates relative to its task's wall-clock budget: a run whose agent runtime reaches 95\% of the budget is a \emph{timeout}, one that ends in the bottom decile of the task's runtimes is an \emph{early stop}, and any other failure is a \emph{wrong solution}; a small residual of provider- and verifier-side errors is excluded.

\section{Experimental Setup}\label{sec:setup}

The evaluated unit in all experiments is a frozen model-harness \emph{bundle}: which defines a model and the agent harness that runs it. The twelve models are GPT-5.6 Sol, GPT-5.6 Terra, and GPT-5.6 Luna (OpenAI); Claude Sonnet 5, Claude Opus 4.8, Claude Fable 5, and Claude Opus 5 (Anthropic); Kimi K3, GLM-5.2, Qwen3.7 Max, and MiniMax M3; and Grok~4.5 (xAI). In the primary comparison, Anthropic models use Claude Code, OpenAI models use Codex, Grok~4.5 uses Grok Build, and the remaining models use Terminus-2. All bundles use high effort and are evaluated on all 89 tasks under every policy configuration, with exactly three valid trials per bundle-task-policy cell.

Claude Fable 5 ships with dual-use safety classifiers, and Anthropic's default serves flagged requests with Claude Opus~4.8 instead; we keep this default, since the gated route is what an organization deploys.

\paragraph{Trial protocol.}
A trial runs the bundle headlessly until the harness returns or the task's timeout is reached; either way the official verifier scores the resulting environment, so a timed-out execution can still pass, and its PASS or FAIL is the trial score. Harness-internal retries are part of the trial. Provisioning, pre-flight, provider, harness, or verifier malfunctions invalidate an attempt, which we log, exclude, and rerun until the cell holds three valid trials; a scored trial is never rerun.

\paragraph{Task-level cost sampling.}
To characterize per-run cost distributions, we picked two tasks and evaluated them with the GPT-5.6 Luna--Codex bundle at 100 runs per condition under control and \textit{NIST-derived high}: \taskname{compile-compcert}, whose success rate hardening leaves nearly unchanged, and \taskname{caffe-cifar-10}, whose success rate degrades sharply. We read all 200 hardened-condition trajectories to attribute the observed cost movement to a mechanism.

\paragraph{Model-harness ablation study.}
To test whether the two cost extremes of the Pareto frontier follow the model or its harness, we evaluate GPT-5.6 Luna, which anchors the low-cost side, and Grok~4.5, which anchors the high-cost side, with both Codex and Grok Build. This experiment disentangles the effect of the harness from the model's performance in the restricted environment.

\paragraph{Infrastructure.}
Each trial runs in an isolated cloud sandbox (Daytona) built from the task's own Terminal-Bench image; the sandbox's CPU and memory follow that task's Terminal-Bench specification, which we do not override. Model inference is served over a single OpenRouter route with per-harness vendor endpoints. Software versions are all in Appendix~E.

\begin{figure*}[!t]
  \centering
  \includegraphics[width=\textwidth]{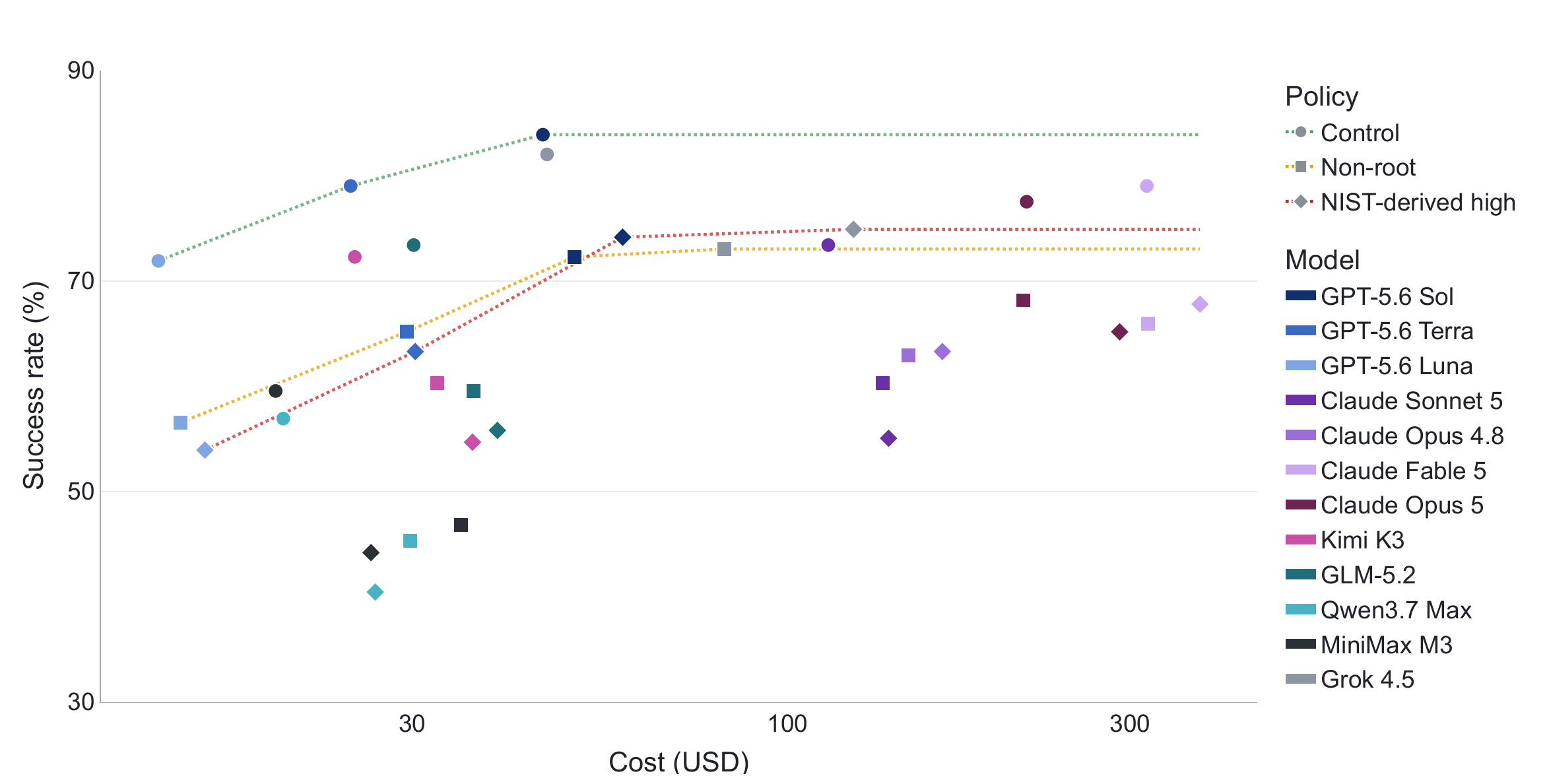}
  \caption{Hardening moves the Pareto frontier down and to the right
  (lower success, higher cost), by amounts that differ across model--harness
  bundles. Each point is one of the 12 bundles under one policy on
  Terminal-Bench 2.1, three trials per task: marker shape encodes the
  policy, color the model. Axes: success rate (\%) against total cost (USD,
  logarithmic scale). Dashed lines trace each policy's Pareto frontier.}
  \label{fig:frontier}
\end{figure*}

\subsection{Robustness Measure}
To measure a bundle's robustness under restriction, we summarize each bundle--policy cell by two numbers, computed per bundle over the task pool: the success rate
$\mathrm{SR}_m(\ell)$, the percentage of passing trials over all tasks and
repetitions of bundle $m$ under condition $\ell$, and the mean cost per task $C_m(\ell)$, in US dollars. A bundle's shift under a restricted condition is the change from its own control values.

\begin{equation}
\label{eq:policy-shift}
\begin{array}{rcl}
\Delta \mathrm{SR}_m(\ell) &=& \mathrm{SR}_m(\ell) -
  \mathrm{SR}_m(\mathrm{control}), \\
\Delta C_m(\ell) &=& \displaystyle 100 \cdot
  \frac{C_m(\ell) - C_m(\mathrm{control})}{C_m(\mathrm{control})},
\end{array}
\end{equation}

For the restricted conditions $\ell \in \{\text{non-root},\allowbreak \text{NIST-derived high}\}$, $\Delta \mathrm{SR}_m(\ell)$ is an absolute difference in percentage points; $\Delta C_m(\ell)$ is normalized by the bundle's own control cost, making it comparable across bundles whose absolute costs differ by orders of magnitude. At task level we use the cost multiplier: a bundle--task pair's mean cost
over completed (PASS or FAIL) runs under a policy divided by its mean
control cost, averaged with equal weight across pairs.

\section{Results}\label{sec:results}

\subsection{The Success--Cost Frontier under Policy}

Figure~\ref{fig:frontier} places each bundle in the success--cost plane under
the three policies. A clear Pareto frontier emerges under every policy, not only
under control: the Codex bundles trace it under control, from the cheapest
(GPT-5.6 Luna) up to the highest-scoring (GPT-5.6 Sol), and under both

hardened policies Grok~4.5 joins its high-cost end; on the 82 tasks with a
solvability witness
Grok~4.5 instead ties with GPT-5.6 Sol under \textit{NIST-derived high} and
leaves the frontier as the costlier of the two (Appendix~F). Tightening the
policy pushes this
frontier lower and further right, trading success for cost. That movement is far
from uniform across bundles: some models forfeit noticeably more task success
under policy than others, and some absorb the policy more as inflated cost than
as lost success. We take up this per-model heterogeneity, and what drives it, in the analyses that follow.

\paragraph{Model--harness ablation.}
Grok~4.5 remains the higher-cost model under both harnesses and all three
policies (Table~\ref{tab:model-harness-ablation}), whereas moving
either endpoint model to the other harness lowers success at every policy.
The Luna--Grok Build bundle moves down and left because its lower spend
accompanies a much larger success loss, not because it offers a more useful
operating point.

\begin{table}[!htb]
  \centering
  \small
  \setlength{\tabcolsep}{2.5pt}
  \begin{tabular}{@{}llrrrr@{}}
    \toprule
    & & \multicolumn{2}{c}{Codex} & \multicolumn{2}{c}{Grok Build} \\
    \cmidrule(lr){3-4}\cmidrule(lr){5-6}
    Policy & Model & $SR$ (\%) & $C$ (\$) & $SR$ (\%) & $C$ (\$) \\
    \midrule
    Control & Luna & 71.9 & 13.28 & 34.5 & 12.85 \\
            & Grok~4.5 & 76.8 & 61.69 & 82.0 & 46.20 \\
    \addlinespace
    Non-root & Luna & 56.6 & 14.25 & 27.3 & 9.04 \\
             & Grok~4.5 & 65.5 & 98.83 & 73.0 & 81.59 \\
    \addlinespace
    NIST-derived & Luna & 53.9 & 15.41 & 25.5 & 10.26 \\
    high & Grok~4.5 & 62.9 & 122.32 & 74.9 & 123.49 \\
    \bottomrule
  \end{tabular}

  \caption{Grok~4.5 remains the higher-cost model under either
  harness at every policy, while both crossed pairings reduce success.
  Success rate ($SR$) and mean 89-task replicate cost ($C$) for the
  $2\times2$ model--harness ablation, with three valid trials per task (267
  trials per policy); trials on blocked-by-design tasks count as
  non-passing.}
  \label{tab:model-harness-ablation}
\end{table}

\subsection{Restriction Sensitivity}
\label{sec:policy-shifts}

\begin{figure}[t]
  \centering
  \includegraphics[width=\columnwidth]{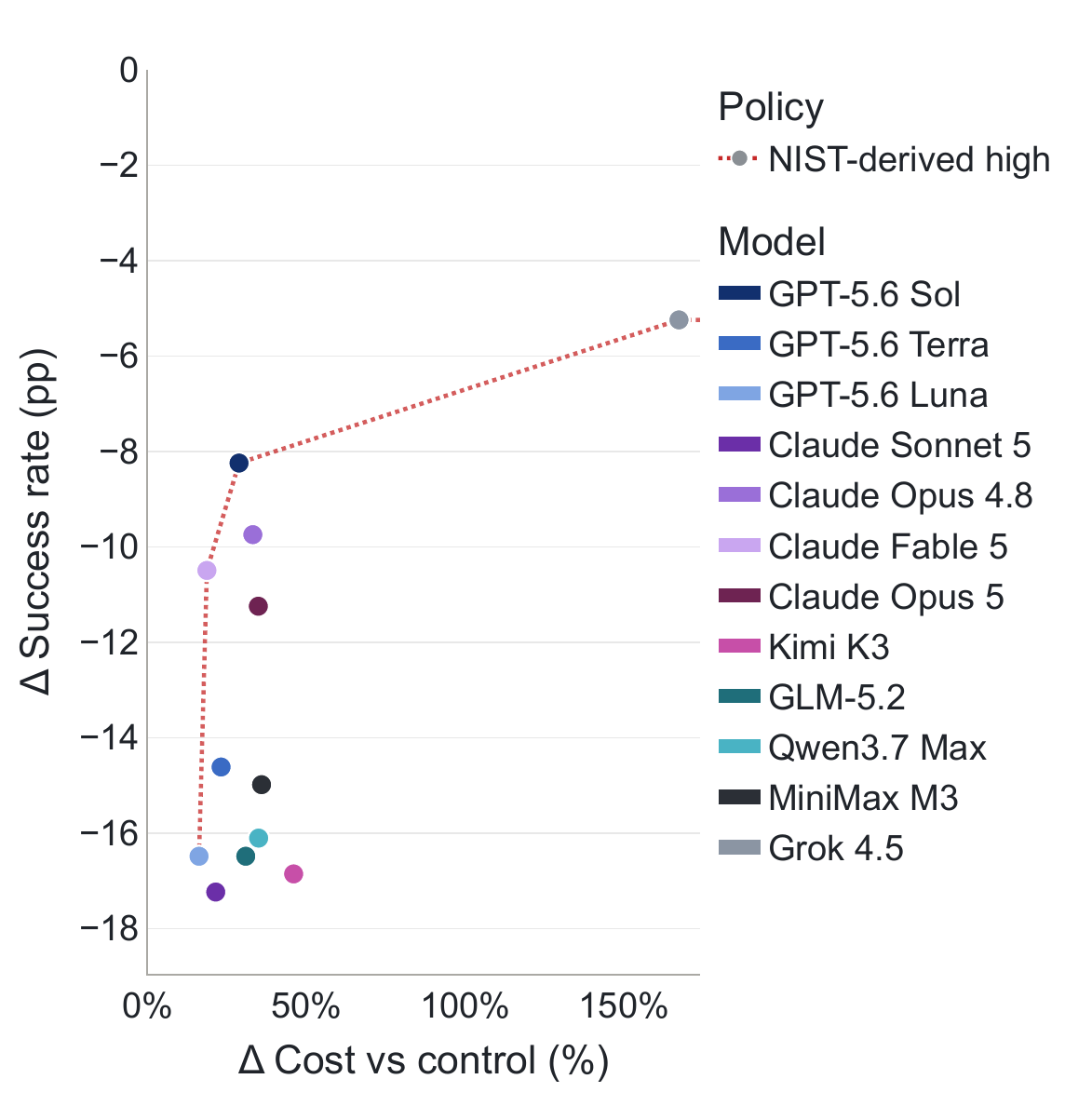}
   \caption{Every bundle loses success and gains cost non-uniformly
  under NIST-derived high. The smallest success loss (Grok 4.5)
  is bought with the largest cost inflation. Each point is one bundle's
  shift from its own control values on the 89-task pool, three trials per
  task (Eq.~(\ref{eq:policy-shift})): the horizontal axis is the change in
  mean cost per task, $\Delta C_m$, in percent of the bundle's control
  cost; the vertical axis the success-rate change, $\Delta \mathrm{SR}_m$,
  in percentage points; color encodes the model. The dotted line traces a
  robustness Pareto frontier: the bundles offering the best trade-off
  between success loss and cost inflation under restriction.}
  \label{fig:policy-shifts}
\end{figure}

Figure~\ref{fig:policy-shifts} plots the shifts of
Eq.~(\ref{eq:policy-shift}), normalized by each bundle's control operating
point in Figure~\ref{fig:frontier}. Each task contributes its three valid
trials, scored by the original verifiers, so these numbers predate the
verifier repairs and the artifact separation of the next subsection applies
on top of them.

Under \textit{NIST-derived high}, every bundle's point estimate worsens on
both axes: hardening lowers the success rate and raises the mean cost per
task for all twelve bundles, so every point in
Figure~\ref{fig:policy-shifts} falls in the same quadrant. Within that
quadrant the two axes do not move together, and bundles pay in different
currencies. Grok 4.5 gives up the least success ($-7.1$~pp) but absorbs the
largest cost inflation by a wide margin ($+167.3\%$); Claude Sonnet 5 loses
the most success ($-18.3$~pp) at a comparatively modest cost inflation
($+21.4\%$); GPT-5.6 Luna is the mirror of Grok 4.5, showing the smallest
cost inflation ($+16.0\%$) with a success loss nearly as large

($-18.0$~pp). With three trials per cell and no intervals we report

this as a spread across bundles, not as a ranking. All five
verifier-repaired tasks lie inside this pool, and their repairs raise
success under \textit{NIST-derived high} on two of them, so the losses reported here
are slightly overstated (next subsection).

\subsection{Benchmark Artifacts under Policy}

Because a sufficiently strict policy can render a task not merely harder but unsolvable, we begin by establishing, for every Terminal-Bench task, whether it remains solvable under \textit{NIST-derived high} enforcement. Most tasks remain solvable: 82 of the 89 have a solvability witness, a trajectory that reaches the official passing state under the policy. For 50 the witness is Terminal-Bench's own reference solution, unchanged; for the other 32 we author a policy-compliant reference solution ourselves, changing no task (Appendix~B). We call these the \emph{unaffected} (50) and \emph{affected} (32) subsets.

Seven tasks are \emph{blocked by design}: no policy-compliant solution exists. This is itself a finding that reinforces our central claim, their requirements are incompatible with a hardened environment, and we keep all 89 tasks in the reported results, analyzing the slices separately.

Wherever a verifier indicated failure although the agent had achieved the
stated goal, the root cause was an \emph{over-specified} verifier: an
assertion demanding an incidental detail that the policy blocks but the
task never required, for example one exact installation path for a tool
that only needs to be available, a failure mode documented in
code-generation benchmarks \citep{sharifloo_where_2025}. For each of the
five such verifiers we additively broaden the assertion to also accept the
agent's writable workspace, leaving every substantive check intact; the
identical verifier runs at every policy level, including control, with
original outcomes kept for audit.

The headline results (Figure~\ref{fig:frontier}) include both artifact
classes. Rescoring under \textit{NIST-derived high} increases the success
rate on two of the five repaired tasks (Table~\ref{tab:verifier-repairs}),
both recoveries with the same shape: the agent had built the required tool
in its writable workspace, where the original check never looked.

The other three are unchanged under either verifier, confirming the repairs
correct the checks rather than inflate scores: one is unsolved even in
control, and two fail for genuine hardening reasons; on
\taskname{mcmc-sampling-stan} every bundle installed the sampler and
computed correct posterior means, yet none made the installation visible to
a second process, so checks that open a fresh session fail in every trial
(Appendix~B).

\begin{table}[!t]
  \centering
  \small
  \setlength{\tabcolsep}{2.5pt}

  \begin{tabular}{lrrrr}
    \toprule
    Task & Orig. & Adapted & $\Delta$ (pp) & Control \\
    \midrule
    sqlite-with-gcov & 0.0\% & 80.6\% & +80.6 & 97.2\% \\
    adaptive-rejection-sampler & 0.0\% & 19.4\% & +19.4 & 61.1\% \\
    configure-git-webserver & 19.4\% & 19.4\% & 0.0 & 58.3\% \\
    mcmc-sampling-stan & 0.0\% & 0.0\% & 0.0 & 88.9\% \\
    make-doom-for-mips & 0.0\% & 0.0\% & 0.0 & 0.0\% \\
    \bottomrule
  \end{tabular}
  \caption{The original verifiers failed to credit valid solutions on
  two of the five tasks. Success rate under NIST-derived high scored
  by the original (Orig.) and by the adapted verifier, and in control with the
  original verifier. Each cell is 36 trials: twelve model--harness bundles,
  three trials each. $\Delta$ is Adapted minus Orig.\ in percentage points; the
  gap between Adapted and Control is the residual policy effect.}
  \label{tab:verifier-repairs}
\end{table}

Under hardening the seven blocked-by-design tasks are guaranteed failures
for every bundle alike: restricting to the 82 witnessed tasks raises each
bundle's hardened success by 3.4 to 5.8 points while leaving control
essentially unchanged, removing a roughly common five to seven points from
every measured penalty. What remains discriminates: the control-to-hardened
drop runs from 2.4 points (Grok 4.5) to 13.5 (Claude Sonnet 5) on the
reduced pool, against 7.1 to 18.3 on the full pool; Appendix~F reports both
pools in full, every bundle's success rate and cost under each policy.

\paragraph{Provider-side safety interventions.}
Anthropic and OpenAI operate safety classifiers inside their serving layers
that can end a trial before the verifier runs; they respond to request
content, and any organization calling the same route encounters them.
Across the campaign they affected five of the twelve bundles and 98 trials. We retain every affected trial
in the reported aggregates, consistent with measuring the bundle as
deployed. Only the OpenAI route retries a dropped stream; every Anthropic
refusal ends its trial. Claude Fable 5's declared Opus~4.8 fallback
(Experimental Setup) served 29.9\% of its requests, so its results describe
the gated route rather than Fable 5 alone. Intervention-caused failures cost any bundle at most 2.2 percentage points of measured success, and incidence does not rise with policy severity; Appendix~D documents each mechanism, message, and count.

\subsection{Mechanisms of Degradation}

On the 82 tasks that admit a solvability witness, hardening still costs 7.4
points of success (72.5\% to 65.1\%, Table~\ref{tab:failure-decomp}).
Classifying each failed run by where it terminates relative to its task's
wall-clock budget shows the additional failures are of two kinds: runs that
exhaust the budget (timeouts, 12.0\% to 16.4\% of runs) and runs that end
well within it with a solution that fails verification (wrong solutions,
12.8\% to 15.5\%). Agents do not give up more often under policy: early
stops barely move (2.4\% to 2.7\%). Of the failures hardening adds, roughly
59\% are timeouts and 37\% wrong solutions.

\begin{table}[!t]
  \centering
  \small

  \setlength{\tabcolsep}{4pt}
  \begin{tabular}{lrrr}
    \toprule
    Outcome (\% of runs) & Control & Non-root & NIST-derived high \\
    \midrule
    Success     & 72.5 & 66.9 & 65.1 \\
    Timeout     & 12.0 & 14.5 & 16.4 \\
    Wrong solution & 12.8 & 15.5 & 15.5 \\
    Early stop  & 2.4 & 2.6 & 2.7 \\
    \bottomrule
  \end{tabular}
  \caption{Hardening converts successes into timeouts and wrong
  solutions, not into early stops. Run outcomes per policy condition on the
  82 solvable-witness tasks (12 bundles, three trials per task, 2{,}952 runs
  per condition). A failed run counts as a timeout when its agent runtime
  reaches 95\% of the task's wall-clock budget, as an early stop when it ends
  in the bottom decile of that task's runtimes, and as a wrong solution
  otherwise; a residual 0.4\% of runs per condition (provider- and
  verifier-side errors) is excluded.}
  \label{tab:failure-decomp}
\end{table}

The same pressure is visible in the runs that succeed: matching passing runs
within the same bundle and task, a hardened pass takes 13\% more wall-clock
time, 14\% more tool calls, and 26\% more tokens than its control
counterpart. Under \textit{NIST-derived high} the timeout class grows fast
enough to overtake wrong solutions as the most common failure. The mix is
bundle-dependent: the open-weight bundles evaluated on Terminus-2 account
for most of the additional timeouts, while the Codex bundles add almost
exclusively wrong solutions within budget.

\subsection{Cost Inflation under Policy}

\begin{figure}[!t]
  \centering
  \includegraphics[width=0.75\columnwidth]{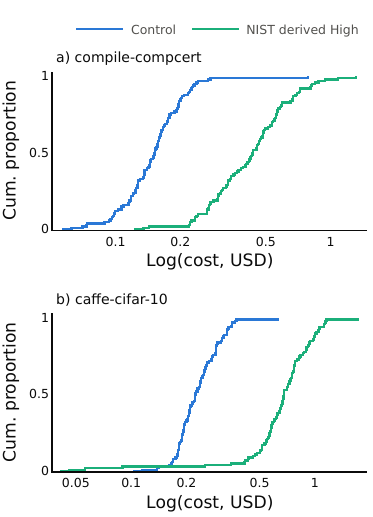}
  \caption{Policy enforcement raises cost even where success holds:
  median per-run cost roughly triples under NIST-derived high on both
  tasks. Empirical cumulative distributions of per-run cost (USD,
  logarithmic scale) for Codex (GPT-5.6 Luna) on two selected
  Terminal-Bench tasks, 100 runs per condition. Success holds on
  \texttt{compile-compcert} (a, 84\%) and degrades on
  \texttt{caffe-cifar-10} (b, 90\% to 29\%); the median per-run cost rises
  $2.9\times$ (a) and $3.0\times$ (b). The median cost of failing runs is
  close to that of passing runs ($0.96\times$ (a), $1.11\times$ (b)), suggesting
  policy-induced failures consume roughly the full inflated budget.}
  \label{fig:cost_cdf}
\end{figure}

Hardening's second currency is cost.
Figure~\ref{fig:cost_cdf} shows how the \textit{NIST-derived high} policy shifts the cost distributions of the two deep-sampled tasks. The median cost increase is comparable in both cases, decoupling the cost of policy from its effect on success. Passing and failing runs pay the shift alike. The mechanism is almost exclusively workaround construction (97\%): the agent abandons the blocked path and rebuilds a tool-chain from source or sources substitute data, and this reconstruction produces the added cost; the few low-cost extremes are runs that stop early once blocked.

\begin{figure}[!t]
    \centering
    \includegraphics[width=\columnwidth]{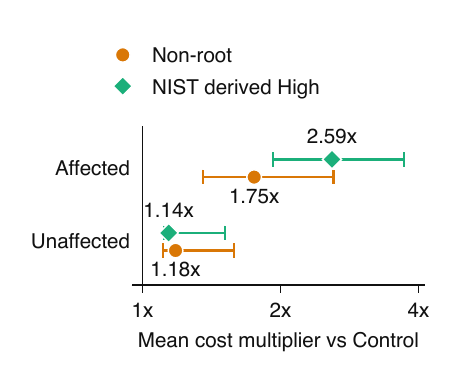}

        \caption{Policy-induced cost inflation concentrates on tasks
          whose canonical reference solutions break under policy. Mean cost
          multipliers relative to control, equal-weighted over the bundle--task
          pairs of all 12 bundles: the affected subset (32 tasks verified only
          by a non-canonical reference solution) rises to $1.75\times$ under
          non-root and $2.59\times$ under NIST-derived high, while the
          unaffected subset (50 tasks whose canonical reference solution passes
          unchanged) stays near parity ($1.18\times$, $1.14\times$). Whiskers:
          95\% intervals from 10{,}000 hierarchical-bootstrap replicates.}
    \label{fig:normalized_cost}
\end{figure}

Figure~\ref{fig:normalized_cost} stratifies the cost effect by

reference-solution compatibility. Adding the seven blocked-by-design tasks
to the affected subset gives 39 tasks and yields

$3.03\times$ [1.81, 5.56] under non-root and $3.48\times$ [2.37, 5.56].
``Affected'' describes reference-solution evidence rather than every agent
trajectory: an agent may follow a policy-affected trajectory on a task with
an unaffected reference solution, consistent with the small residual
inflation in the unaffected subset, or, less often, an unaffected
trajectory on an affected-label task. The intervals quantify within-subset
uncertainty rather than testing between-subset differences.

The same stratification predicts the success penalty. On the affected subset
the success rate falls from 70.3\% to 54.5\% under \textit{NIST-derived
high} (15.8 points), while on the unaffected subset it moves from 73.8\% to
71.8\%, a change indistinguishable from noise; 83.5\% of the additional
failures in Table~\ref{tab:failure-decomp} fall on the 32 affected tasks,
and the trajectory lengthening concentrates there as well (36\% more
wall-clock time and 70\% more tokens for matched passing runs, against no
measurable change on unaffected tasks). The two subsets have nearly
identical control success rates, so the divergence is not a difficulty
difference between the groups: it appears only once the policy is enforced.
A task's reference-solution compatibility, computable before any agent
runs, therefore anticipates where a policy will cost both money and
success.

\subsection{Blocked Actions under Policy}
\label{sec:blocked-actions}

Finally, we measure the restrictions agents actually hit. A \emph{blocked
action} is one denied operation, counted from verified-complete evidence
only (Appendix~G). By design, control has no blocked actions; under
\textit{non-root} every blocked action is a privilege denial. Under
\textit{NIST-derived high}, a third of trials (675 of 2{,}051) record at
least one. Egress denials lead on every measure (500 trials, against 283
privilege and 158 filesystem), and their 90.5\% share of volume is mostly
persistence: a trial that hits the egress wall records 10.5 denials on
average, against 1.4 for privilege. Blocks concentrate sharply across the 88 tasks with verified evidence
(the remaining task's traces are always truncated): 14 record a verified zero, 48 record five or fewer in total, and five
carry half the volume (\taskname{build-pov-ray},
\taskname{break-filter-js-from-html}, \taskname{mteb-retrieve},
\taskname{protein-assembly}, \taskname{count-dataset-tokens}). 

Exposure follows solvability: trials with at least one blocked action rise
from 14.4\% on the unaffected subset, to 51.6\% on the affected subset, to
67.8\% on tasks blocked by design, and on affected tasks failing trials
record 2.7 times the blocked actions of passing ones. These are diagnostic
associations, not causes (Appendix~G).

\section{Limitations}

\textbf{Single benchmark.} All evidence comes from Terminal-Bench and its
89 tasks. Boundary-Bench attaches to the runtime rather than to task content,
but every added benchmark would repeat the solvability audit, the witness
authoring, and the inference spend of the full twelve-bundle grid; our
findings are therefore established for this benchmark only.

\textbf{Bundle-level attribution.} The evaluated unit is the frozen
model--harness bundle, so effects are not attributed to model weights or the
harness alone. The ablation of Table~\ref{tab:model-harness-ablation} covers
only the frontier's two cost extremes, and a harness change alters the
prompt, tools, and interface; full-roster attribution remains bundle-level.

\textbf{One operating point, three trials.} Every primary bundle runs at high
reasoning effort with three valid trials per bundle--task--policy cell; no
other effort setting or trial count is evaluated. Intervals describe
run-to-run variability at this configuration; Pareto, sensitivity, and
ordering statements are descriptive, and small differences between bundles
should not be over-read.

\textbf{Policy realism.} The three conditions form one severity ladder derived
from common enterprise restriction settings and mapped to NIST SP 800-53
families. They are not a certified deployment configuration, and we measure
their performance cost, not their effectiveness as security.

\section{Conclusion}

Security restrictions are part of the coding agent's real operating
environment, yet benchmarks measure agents without them. Evaluating twelve model--harness bundles on Terminal-Bench 2.1
under three natively enforced policy levels, we draw four conclusions.

\textbf{Hardening is non-uniform.} Every bundle loses success and gains
cost under the strictest policy, \textit{NIST-derived high}, but by amounts that differ across bundles:
success losses reach 18.3 points and cost inflation 167.3\%.

\textbf{The trade-off is universal; its currency is not.} Some bundles
absorb the policy as inflated cost, others as lost success, and none
escapes both. The loss is grinding rather than surrender: longer
trajectories ending in timeouts or wrong solutions, almost never in early
stops, so budgets must be provisioned under the deployment's policy.

\textbf{The tax is predictable.} Given a policy, reference-solution
compatibility identifies the affected tasks before any agent runs, for
success and cost alike, and the procedure applies to any benchmark that
ships reference solutions.

\textbf{Benchmarks should carry a policy axis.} Hardening degrades even
frontier bundles, restoring discriminative headroom, and model selection is
only valid under the deployment's own policy; reporting performance in a
common restricted environment should become standard practice.

Natural next steps are extending the solvability audit to further benchmarks
and building policy-adaptive agents that mitigate the mechanisms quantified
here.

\section{Acknowledgments}
We thank our colleagues at Accomplish AI for feedback and support
throughout this work. The evaluation campaign, including all model
inference and cloud-sandbox compute, was funded by Accomplish AI.

\bibliography{references}

\clearpage
\section{Appendix A: Policy Construction Details}
\label{app:policy-construction}

This appendix gives the full construction of the \textit{NIST-derived high} policy summarized in the Boundary-Bench section of the main paper: its three hardening axes (network, filesystem, and privilege), the native enforcement and NIST controls each adds, and the pre-flight probes that verify them.

\subsection{Axis Definitions}

Each axis is enforced by a distinct native mechanism, mapped to the NIST
controls it projects onto the runtime. The mapping records the controls that motivated each mechanism and is not a compliance claim. Table~\ref{tab:nfp-axes} summarizes the
three axes at the \textit{NIST-derived high} posture; the paragraphs that follow
give the read-only trees and privilege lockdown, and
Table~\ref{tab:egress-allowlist} reproduces the exact egress allowlist.
\begin{table*}[t]
\centering
\scriptsize
\begin{tabular}{p{0.07\textwidth} p{0.55\textwidth} p{0.16\textwidth}}
\toprule
Axis & Enforcement & Mapped NIST controls \\
\midrule
Network&
All egress is forced through a loopback proxy; a firewall rule scoped to the agent's user rejects any traffic that bypasses it, including from child processes. Cloud-metadata and private-address ranges are always denied. The proxy default-denies every domain outside a fixed 205-domain developer allowlist adapted from the default allowlist for Claude Code on the web, which includes the model-serving endpoints; no task-specific domains are added. A blocked domain returns a bare \texttt{HTTP 403}; direct egress fails with a refused connection.&
SC-7, SC-7(5), SC-7(8) \\
\addlinespace
Filesystem&
Only the
workspace, temporary and shared-memory paths (\texttt{/tmp}, \texttt{/dev/shm}), and the harness runtime and cache directories remain
writable. The home directory, \texttt{/var/tmp}, and the operating-system trees are bind-remounted read-only, so writes fail with \texttt{EROFS}.&
AC-3, SC-39, AC-4, SC-4, AC-4(4), SC-7(21), CM-5(1) \\
\addlinespace
Privilege&
The agent runs as a dedicated ordinary user. It has no sudoers entry, and the
root password is locked, so \texttt{sudo} and \texttt{su} escalation fail. The kernel's \texttt{no\_new\_privs} flag is set, the capability bounding set is
empty, and setuid bits are removed. Privileged kernel operations fail with
\texttt{EPERM}.&
AC-6, AC-6(2), AC-6(3), AC-6(5), AC-6(10), CM-5(1)\\
\bottomrule
\end{tabular}
\caption{\textbf{Every restriction is a native operating-system mechanism rather than a policy shim, mapped to the NIST SP 800-53 controls it projects onto the runtime; the agent observes only ordinary system errors (\texttt{HTTP 403}, \texttt{EROFS}, \texttt{EPERM}).} The three runtime-hardening axes, network ($N$), filesystem ($F$), and privilege ($P$), under the \textit{NIST-derived high} policy. The model-serving endpoints are the OpenRouter domains (\texttt{api.openrouter.ai}, \texttt{openrouter.ai}, \texttt{*.openrouter.ai}); the running harness's own vendor endpoint (e.g., \texttt{api.anthropic.com} for Claude Code, \texttt{chatgpt.com} for Codex) is admitted at runtime. The harness runtime and cache directories are \texttt{\textasciitilde/.cache}, \texttt{\textasciitilde/.config}, \texttt{\textasciitilde/.local}, and the running harness's own directory (e.g., \texttt{\textasciitilde/.claude}, \texttt{\textasciitilde/.codex}).}
\label{tab:nfp-axes}
\end{table*}

\paragraph{Network implementation.}
Under \textit{NIST-derived high}, all egress is forced through a loopback proxy that
default-denies every domain outside its allowlist: a fixed 205-domain developer
list adapted from the default allowlist for Claude Code on the
web,\footnote{\url{https://code.claude.com/docs/en/claude-code-on-the-web\#default-allowed-domains}} which includes the model-serving endpoints, plus the running harness's own
vendor endpoint; the full list is reproduced in
Table~\ref{tab:egress-allowlist}. No task-specific domains are added: we extracted each task's
network dependencies from its official reference solution (36 of the 89 tasks
require at least one, spanning 30 distinct domains, of which 9 already fall
inside the allowlist), and deliberately did not admit the remaining 21, so the
allowlist is not fitted to the evaluated tasks. Exactly one task is thereby
foreclosed by an egress blockage alone (\texttt{count-dataset-tokens};
Appendix~B): for every other out-of-list dependency,
an equivalent artifact is reachable through an allowlisted host, so a solver can
route around the blockage, for example, installing the same R packages from the
allowlisted conda-forge channel when CRAN is blocked, or obtaining a toolchain
from the allowlisted Ubuntu mirrors rather than Debian's. The proxy runs with
root ownership while the firewall rule is scoped to the agent's user, so the
proxy cannot be bypassed or disabled from inside the policy; the same rules deny
the cloud-metadata endpoint and private address ranges, and resolved addresses
are rechecked so an allowed domain cannot resolve into a denied range. A
blocked domain returns a bare \texttt{HTTP 403}; direct egress that skips the
proxy fails with a refused connection.

\paragraph{Filesystem implementation.}
The protected operating-system trees are \texttt{/usr}, \texttt{/bin},
\texttt{/sbin}, \texttt{/lib}, \texttt{/lib64}, \texttt{/etc},
\texttt{/opt}, \texttt{/boot}, \texttt{/srv}, \texttt{/root},
\texttt{/var}, and \texttt{/home}; each is bind-remounted read-only. The
working directories (\texttt{/app}, \texttt{/workspace}), the scratch paths
(\texttt{/tmp}, \texttt{/dev/shm}), and the agent's runtime and cache
directories (\texttt{\textasciitilde/.cache}, \texttt{\textasciitilde/.config},
\texttt{\textasciitilde/.local}, and the running harness's own directory such
as \texttt{\textasciitilde/.claude} or \texttt{\textasciitilde/.codex}) remain
writable; the rest of the home directory and \texttt{/var/tmp} are read-only.
Writes outside the writable set fail with \texttt{EROFS}. Reads remain available everywhere; the mechanism restricts tampering rather than confidentiality.

\paragraph{Privilege implementation.}
Under both \textit{non-root} and \textit{NIST-derived high} restricted policies, \texttt{setpriv} drops the agent to a dedicated user and group; the user has no sudoers entry, and the root password
is locked, so \texttt{sudo} and \texttt{su} escalation fail. Both policies also install a
firewall rule, scoped to the agent's user and outside the network policy axis,
that rejects access to the sandbox provider's own control-plane ports, closing
a platform escalation path.
\textit{NIST-derived high} additionally sets \texttt{no\_new\_privs} so no executed program can gain
privilege, empties the Linux capability bounding set, and strips \texttt{setuid} bits
from the task image. Mounting filesystems, changing firewall rules, and
opening raw sockets then fail with \texttt{EPERM}.

\subsection{Quality Control}
Enforcement is checked by a pre-flight probe suite, run once per sandbox
before the agent starts and outside task-execution time, that exercises every
configured blockage: \emph{direct} probes confirm each restriction denies its target operation,
\emph{adversarial} probes attempt bypasses (nested shells, raw-IP and
alternate-port egress, child-process inheritance), and \emph{enablement} probes
confirm permitted operations still succeed. Because these probes run before the agent starts, a required probe failure can never be classified as an agent failure or scored against the task. Instead, the attempt is discarded as an infrastructure error and replaced by a fresh sandbox.
A hardening step that cannot be established likewise fails closed rather than
running unhardened. The probes leave no trace; their scratch area is verifiably removed before the agent begins.

\onecolumn
\begingroup\scriptsize\setlength{\tabcolsep}{3pt}
\begin{longtable}{@{}lll@{}}
\caption{\textbf{The fixed 205-domain egress allowlist of \textit{NIST-derived
high}.} All egress passes a default-deny proxy admitting only these domains.
A bare entry matches its exact host only; a \texttt{*.}-prefixed entry matches
strict subdomains at any depth, so \texttt{openrouter.ai} and
\texttt{*.openrouter.ai} are distinct entries. The running harness's own
vendor endpoint is additionally admitted at runtime, and no task-specific
domains are added.}\label{tab:egress-allowlist}\\
\toprule
\endfirsthead
\toprule
\endhead
\bottomrule
\endfoot
\texttt{api.openrouter.ai} & \texttt{openrouter.ai} & \texttt{*.openrouter.ai} \\
\texttt{api.anthropic.com} & \texttt{statsig.anthropic.com} & \texttt{docs.claude.com} \\
\texttt{platform.claude.com} & \texttt{code.claude.com} & \texttt{claude.ai} \\
\texttt{github.com} & \texttt{www.github.com} & \texttt{api.github.com} \\
\texttt{npm.pkg.github.com} & \texttt{raw.githubusercontent.com} & \texttt{pkg-npm.githubusercontent.com} \\
\texttt{objects.githubusercontent.com} & \texttt{release-assets.githubusercontent.com} & \texttt{codeload.github.com} \\
\texttt{avatars.githubusercontent.com} & \texttt{camo.githubusercontent.com} & \texttt{gist.github.com} \\
\texttt{gitlab.com} & \texttt{www.gitlab.com} & \texttt{registry.gitlab.com} \\
\texttt{bitbucket.org} & \texttt{www.bitbucket.org} & \texttt{api.bitbucket.org} \\
\texttt{registry-1.docker.io} & \texttt{auth.docker.io} & \texttt{index.docker.io} \\
\texttt{hub.docker.com} & \texttt{www.docker.com} & \texttt{production.cloudflare.docker.com} \\
\texttt{download.docker.com} & \texttt{gcr.io} & \texttt{*.gcr.io} \\
\texttt{ghcr.io} & \texttt{mcr.microsoft.com} & \texttt{*.data.mcr.microsoft.com} \\
\texttt{public.ecr.aws} & \texttt{cloud.google.com} & \texttt{accounts.google.com} \\
\texttt{gcloud.google.com} & \texttt{*.googleapis.com} & \texttt{storage.googleapis.com} \\
\texttt{compute.googleapis.com} & \texttt{container.googleapis.com} & \texttt{azure.com} \\
\texttt{portal.azure.com} & \texttt{microsoft.com} & \texttt{www.microsoft.com} \\
\texttt{*.microsoftonline.com} & \texttt{packages.microsoft.com} & \texttt{dotnet.microsoft.com} \\
\texttt{dot.net} & \texttt{visualstudio.com} & \texttt{dev.azure.com} \\
\texttt{*.amazonaws.com} & \texttt{*.api.aws} & \texttt{oracle.com} \\
\texttt{www.oracle.com} & \texttt{java.com} & \texttt{www.java.com} \\
\texttt{java.net} & \texttt{www.java.net} & \texttt{download.oracle.com} \\
\texttt{yum.oracle.com} & \texttt{registry.npmjs.org} & \texttt{www.npmjs.com} \\
\texttt{www.npmjs.org} & \texttt{npmjs.com} & \texttt{npmjs.org} \\
\texttt{yarnpkg.com} & \texttt{registry.yarnpkg.com} & \texttt{pypi.org} \\
\texttt{www.pypi.org} & \texttt{files.pythonhosted.org} & \texttt{pythonhosted.org} \\
\texttt{test.pypi.org} & \texttt{pypi.python.org} & \texttt{pypa.io} \\
\texttt{www.pypa.io} & \texttt{rubygems.org} & \texttt{www.rubygems.org} \\
\texttt{api.rubygems.org} & \texttt{index.rubygems.org} & \texttt{ruby-lang.org} \\
\texttt{www.ruby-lang.org} & \texttt{rubyforge.org} & \texttt{www.rubyforge.org} \\
\texttt{rubyonrails.org} & \texttt{www.rubyonrails.org} & \texttt{rvm.io} \\
\texttt{get.rvm.io} & \texttt{crates.io} & \texttt{www.crates.io} \\
\texttt{index.crates.io} & \texttt{static.crates.io} & \texttt{rustup.rs} \\
\texttt{static.rust-lang.org} & \texttt{www.rust-lang.org} & \texttt{proxy.golang.org} \\
\texttt{sum.golang.org} & \texttt{index.golang.org} & \texttt{golang.org} \\
\texttt{www.golang.org} & \texttt{goproxy.io} & \texttt{pkg.go.dev} \\
\texttt{maven.org} & \texttt{repo.maven.org} & \texttt{central.maven.org} \\
\texttt{repo1.maven.org} & \texttt{repo.maven.apache.org} & \texttt{jcenter.bintray.com} \\
\texttt{gradle.org} & \texttt{www.gradle.org} & \texttt{services.gradle.org} \\
\texttt{plugins.gradle.org} & \texttt{kotlinlang.org} & \texttt{www.kotlinlang.org} \\
\texttt{spring.io} & \texttt{repo.spring.io} & \texttt{packagist.org} \\
\texttt{www.packagist.org} & \texttt{repo.packagist.org} & \texttt{nuget.org} \\
\texttt{www.nuget.org} & \texttt{api.nuget.org} & \texttt{pub.dev} \\
\texttt{api.pub.dev} & \texttt{hex.pm} & \texttt{www.hex.pm} \\
\texttt{cpan.org} & \texttt{www.cpan.org} & \texttt{metacpan.org} \\
\texttt{www.metacpan.org} & \texttt{api.metacpan.org} & \texttt{cocoapods.org} \\
\texttt{www.cocoapods.org} & \texttt{cdn.cocoapods.org} & \texttt{haskell.org} \\
\texttt{www.haskell.org} & \texttt{hackage.haskell.org} & \texttt{swift.org} \\
\texttt{www.swift.org} & \texttt{archive.ubuntu.com} & \texttt{security.ubuntu.com} \\
\texttt{ubuntu.com} & \texttt{www.ubuntu.com} & \texttt{*.ubuntu.com} \\
\texttt{ppa.launchpad.net} & \texttt{launchpad.net} & \texttt{www.launchpad.net} \\
\texttt{*.nixos.org} & \texttt{dl.k8s.io} & \texttt{pkgs.k8s.io} \\
\texttt{k8s.io} & \texttt{www.k8s.io} & \texttt{releases.hashicorp.com} \\
\texttt{apt.releases.hashicorp.com} & \texttt{rpm.releases.hashicorp.com} & \texttt{archive.releases.hashicorp.com} \\
\texttt{hashicorp.com} & \texttt{www.hashicorp.com} & \texttt{repo.anaconda.com} \\
\texttt{conda.anaconda.org} & \texttt{anaconda.org} & \texttt{www.anaconda.com} \\
\texttt{anaconda.com} & \texttt{continuum.io} & \texttt{apache.org} \\
\texttt{www.apache.org} & \texttt{archive.apache.org} & \texttt{downloads.apache.org} \\
\texttt{eclipse.org} & \texttt{www.eclipse.org} & \texttt{download.eclipse.org} \\
\texttt{nodejs.org} & \texttt{www.nodejs.org} & \texttt{developer.apple.com} \\
\texttt{developer.android.com} & \texttt{pkg.stainless.com} & \texttt{binaries.prisma.sh} \\
\texttt{statsig.com} & \texttt{www.statsig.com} & \texttt{api.statsig.com} \\
\texttt{sentry.io} & \texttt{*.sentry.io} & \texttt{downloads.sentry-cdn.com} \\
\texttt{http-intake.logs.datadoghq.com} & \texttt{browser-intake-us5-datadoghq.com} & \texttt{*.datadoghq.com} \\
\texttt{*.datadoghq.eu} & \texttt{api.honeycomb.io} & \texttt{sourceforge.net} \\
\texttt{*.sourceforge.net} & \texttt{packagecloud.io} & \texttt{*.packagecloud.io} \\
\texttt{fonts.googleapis.com} & \texttt{fonts.gstatic.com} & \texttt{json-schema.org} \\
\texttt{www.json-schema.org} & \texttt{json.schemastore.org} & \texttt{www.schemastore.org} \\
\texttt{*.modelcontextprotocol.io} & & \\
\end{longtable}
\endgroup
\twocolumn

\section{Appendix B: Solvability Witnesses under NIST-derived high}
\label{app:refsol-feasibility}

Before attributing a model's failure to the model, we require a
\emph{solvability witness} for the task: at least one solution trajectory that
reaches the official passing state under the policy. Each task's official
reference solution is the natural first candidate---a \emph{compatibility probe}. We therefore replayed
every task's shipped reference solution under \textit{NIST-derived high} (a fixed trusted 205-domain developer egress allowlist
with no per-task domains, read-only operating-system and home trees, and
non-root execution with capabilities dropped), and treated a task as compatible
only when some trajectory reached the official passing state.

\paragraph{Adapting reference solutions to the policy.}
Many shipped solutions assume the unrestricted control environment: they install
packages system-wide with \texttt{apt-get install}, escalate with \texttt{sudo},
or fetch from hosts outside the allowlist. Under \textit{NIST-derived high} these steps fail
with \texttt{EROFS}, \texttt{EPERM}, or a blocked-egress error. For such tasks we
authored solvability witnesses that reach the same passing state
using only operations the policy permits: rootless installation into the
writable workspace (a fully pinned (\texttt{@EXPLICIT}) conda-forge transaction, or a
download-and-extract \texttt{apt} sysroot pattern), with all fetches restricted
to the trusted allowlist. A witness changes neither the task environment nor its
verifier---it is one admissible trajectory, which is exactly what the
compatibility probe requires.

\paragraph{Diagnosing residual infeasibility.}
A residual set of tasks admitted no witness. Examining these one by one, we found
two distinct causes that call for opposite responses.

The first is a genuine \emph{task--policy conflict}: the task's own instruction
mandates a resource or location the policy denies. Six tasks pin a write to a
read-only tree---\taskname{build-pmars} and \taskname{build-pov-ray} instruct the
solver to install a binary to \texttt{/usr/local/bin}; \taskname{build-cython-ext}
and \taskname{kv-store-grpc} to install Python packages into the system-wide
environment; and \taskname{nginx-request-logging} and \taskname{mailman} to place
configuration and logs under \texttt{/etc} and \texttt{/var}. A seventh,
\taskname{count-dataset-tokens}, is foreclosed by the network policy alone: it
requires the exact metadata file of a Hugging Face dataset, reachable only
through Hugging Face's own endpoints (\texttt{huggingface.co} and its
content-delivery hosts), none of which are in the allowlist. The download URL is
minted only by the blocked \texttt{huggingface.co} hop, no allowlisted mirror
exists, and the file cannot be re-derived deterministically; admitting the task
would require exactly the per-task egress exception the fixed-allowlist policy
rejects. Because the
requirement is written into the task, no admissible trajectory exists without
editing the task itself, which we decline to do; these tasks are reported as
unsolved under \textit{NIST-derived high} rather than made to pass.

The second cause is \emph{verifier over-specification}: the instruction is
satisfiable without the denied privilege, but the verifier inspects only the
conventional root-owned location and so marks a legitimate non-root solution as
failing. \texttt{sqlite-with-gcov} is the illustrative case. The task asks the
solver to build SQLite and make it available on the \texttt{PATH}; a non-root
agent does exactly that by building under the writable workspace and exposing the
binary on its own \texttt{PATH}, yet the verifier resolves \texttt{sqlite3} only
through the root user's \texttt{PATH} and reports failure. Since the requested
work was genuinely performed, this is a deficiency of the check, not of the
solution.

\paragraph{Additive verifier corrections.}
For the over-specified cases our operating rule is deliberately asymmetric: we
never alter a task's instruction, but we may correct a verifier that fails to
recognize a valid solution, provided the correction only \emph{adds} an
admissible location and preserves every substantive check. The corrected verifier
resolves the required artifact at the conventional location \emph{or} the
agent-writable workspace, and is applied uniformly across all policy levels so
that no arm gains an advantage. We produced such additive verifiers for five
tasks: \taskname{sqlite-with-gcov}, \taskname{adaptive-rejection-sampler},
\taskname{make-doom-for-mips}, \taskname{mcmc-sampling-stan}, and
\taskname{configure-git-webserver}.

\begin{table}[t]
\centering
\footnotesize
\begin{tabular}{@{}>{\raggedright\arraybackslash}p{0.34\columnwidth}>{\raggedright\arraybackslash}p{0.58\columnwidth}@{}}
\toprule
\textbf{Task} & \textbf{Over-specification $\rightarrow$ additive repair} \\
\midrule
\taskname{sqlite-with-gcov} & Binary resolved via \texttt{PATH} alone; the prompt-pinned \texttt{/app/sqlite} build was never checked $\rightarrow$ also accept \texttt{/app/sqlite} (\texttt{PATH} check retained). \\
\addlinespace
\taskname{adaptive-rejection-sampler} & Bare-name \texttt{Rscript}, resolvable only from a root-writable \texttt{PATH} directory $\rightarrow$ \texttt{PATH}, else the agent's \texttt{/app} R build. \\
\addlinespace
\taskname{make-doom-for-mips} & \texttt{node /app/vm.js} run with an unset working directory, so the binary load depended on the verifier's cwd $\rightarrow$ pin \texttt{cwd=/app}. \\
\addlinespace
\taskname{mcmc-sampling-stan} & Bare-name \texttt{R}/\texttt{Rscript} resolved to the image's baked, RStan-less \texttt{/usr/bin/R} $\rightarrow$ prefer the agent's \texttt{/app} R wrapper, else \texttt{PATH}. \\
\addlinespace
\taskname{configure-git-webserver} & Bare-name \texttt{git}/\texttt{ssh}/\texttt{ssh-keygen}, requiring a system-wide root install $\rightarrow$ resolve \texttt{PATH}-first with \texttt{/app} fallback; git's SSH transport routed through the resolved \texttt{ssh}. \\
\bottomrule
\end{tabular}
\caption{\textbf{Each repair only generalizes a hard-coded root-owned resolution
mechanism (a \texttt{PATH} entry or working directory); every substantive
assertion is untouched, and the identical repaired verifier runs at every policy
condition, including control.} The five over-specified verifiers and their
additive repairs, for tools the task statements only required to function.
Original-verifier outcomes are preserved for audit, and full task-statement
quotes with per-assertion details accompany the released verifier diffs.}
\label{tab:verifier-repair-details}
\end{table}

\paragraph{Outcome.}
After this analysis, \textbf{82 of the 89 tasks} are established to admit a
solvability witness under \textit{NIST-derived high}: for 50 it is the unmodified reference
solution, and for 32 an authored policy-compatible one (Table~\ref{tab:adjusted-refsols}), five of which reach
the passing state only once the over-specified verifier is corrected as above.
The remaining seven are the six task--policy conflicts and
\texttt{count-dataset-tokens}, whose scoring artifact is documented in
Appendix~C.
The distinction matters for interpreting the headline numbers: a task counted as
unsolved under \textit{NIST-derived high} is one the policy genuinely forecloses by the
task's own terms, not one an adequate verifier would have credited.

\subsection{Why solvable tasks still fail under policy}

Five tasks needed an adapted verifier to admit a policy-compliant solution.
Their outcomes under \textit{NIST-derived high}
(Table~2 of the main paper) differ sharply, and the reasons are
instructive.

\paragraph{What the policy removed.}
Hardening does not add friction to a fixed plan; it removes an assumption the
task was built on, and each axis removes a different one. On
\taskname{sqlite-with-gcov} it removes a \emph{destination}: the agent compiles
SQLite without difficulty, but the conventional install path is read-only. On
\taskname{mcmc-sampling-stan} it removes \emph{supply}: every R package channel
the agents reach lies outside the allowlist, and in 31 of the 36 trials the
request to CRAN's primary mirrors returns \texttt{HTTP 403}, after which agents
work down through further mirrors until they find that an allowlisted
operating-system repository happens to carry the package. On
\taskname{configure-git-webserver} it removes the \emph{architecture} the task
presumes: the provided reference trajectory installs system packages, creates a
UNIX account, and runs a privileged-port SSH daemon as root, none of which
survive hardening. Our witness reaches the passing state only by rebuilding
that stack inside the workspace, including an SSH service that authenticates
the expected login without any corresponding system account.

\paragraph{What a repair can recover.}
A repair recovers only what the agent left somewhere predictable. On
\taskname{sqlite-with-gcov} the task statement itself names the build location,
so the agent and the verifier look in the same place and the repair recovers most
of the score (80.6\% under the adapted verifier against 97.2\% in control). Where no location is prescribed, agents
improvise separately and the repair finds little.
\taskname{mcmc-sampling-stan} is the clearest case: the verifier's own checks
on the estimated posterior means pass in all but two trials, so the statistical
work was done, while both checks that open a fresh R session fail in every
trial, because each agent pointed its own script at its own package directory
instead of making the installation discoverable to another process. Our witness
passes by writing a small wrapper that sets the library path for any caller.
Under hardening these agents completed the requested work but left the result
private to themselves.

\paragraph{What is not a policy effect.}
On \taskname{configure-git-webserver} the adapted verifier recovers none of the
failing trials, which is itself evidence that they are not discovery failures.
\taskname{make-doom-for-mips} is unsolved in control as well, so its zero under
policy carries no information about hardening.

\onecolumn
\begin{longtable}{@{}p{2.9cm} c >{\raggedright\arraybackslash}p{4.8cm} >{\raggedright\arraybackslash}p{6.4cm}@{}}
\caption{\textbf{The 32 solvability witnesses.} Each is a study-authored,
policy-compliant reference solution that reaches the official passing state
under \textit{NIST-derived high} where the shipped one fails. Axis: which
policy axes foreclose the official reference (N network, F filesystem,
P privilege); combined entries list every foreclosing axis, and the
separators carry no meaning beyond conjunction. None of the witnesses
signals artifact locations to the
verifier; discovery of the non-root agent's \texttt{/app} artifacts is handled
by the additive adapted verifiers, applied uniformly across every policy
condition.}
\label{tab:adjusted-refsols}\\
\toprule
\textbf{Task} & \textbf{Axis} & \textbf{Official reference failure} & \textbf{Solvability witness} \\
\midrule
\endfirsthead
\toprule
\textbf{Task} & \textbf{Axis} & \textbf{Official reference failure} & \textbf{Solvability witness} \\
\midrule
\endhead
\bottomrule
\endfoot
adaptive-rejection-sampler & F/P & \texttt{apt} installs R system-wide (root). & Rootless apt-sysroot \texttt{r-base-core}+\texttt{openssl} into \texttt{/app/R/root}; relocate R under \texttt{/app} (binary-patch \texttt{Rscript}'s embedded \texttt{R\_HOME} + LD wrapper); \texttt{openssl}-decrypt the protected \texttt{ars.R} with the oracle password; inject the required input-validation, log-concavity check, and \texttt{test()}/sample outputs. Adapted verifier resolves \texttt{/app/R/bin/Rscript}. \\
\addlinespace
sqlite-with-gcov & F/P & \texttt{apt} build tools; installs \texttt{/usr/local/bin/sqlite3} (root). & Rootless apt-sysroot (\texttt{apt-get --download-only build-essential jimsh tclsh}; \texttt{dpkg-deb -x} into \texttt{/app/.rootless}); build SQLite+gcov from the pre-vendored source tarball into \texttt{/app/sqlite}. Adapted verifier resolves \texttt{/app/sqlite/sqlite3}. \\
\addlinespace
bn-fit-modify & N+F/P & R-installs \texttt{bnlearn} 4.9 from CRAN (off-list), root. & Fetch identical \texttt{bnlearn} 4.9 source from the GitHub CRAN mirror; compile into a user R library in \texttt{/app} (zero non-base deps). \\
\addlinespace
caffe-cifar-10 & N+F/P & Native deps + global links + CIFAR-10 from a non-trusted host. & Rootless private-sysroot Caffe build; SHA-pinned canonical CIFAR-10 \texttt{.bin} batches from \texttt{raw.githubusercontent.com}. Verifier already \texttt{/app}-native. \\
\addlinespace
chess-best-move & F/P & \texttt{apt} stockfish + \texttt{pip --break-system-packages} into \texttt{/usr}. & \texttt{pip --target=/app/pylib} numpy+python-chess; drop stockfish, use \texttt{is\_checkmate()} mate-in-one scan; CV image$\to$FEN kept verbatim. \\
\addlinespace
cobol-modernization & N & \texttt{apt} gnucobol3 from Debian mirror (off-list). & Skip \texttt{apt} (gnucobol baked, never used by verifier); emit pure-stdlib port with byte-identical fixed-width output. Zero egress. \\
\addlinespace
compile-compcert & N+F/P & \texttt{opam}/coq-released repos (off-list) + OCaml from source, root. & Rootless apt-sysroot OCaml toolchain from noble universe; build pinned Coq 8.16.1 + CompCert 3.13.1 from GitHub, all in \texttt{/app}. \\
\addlinespace
configure-git-webserver & F/P & \texttt{apt} git/nginx/sshd, \texttt{adduser}, system sshd on \texttt{:22} (root). & conda-forge git+openssh+paramiko into \texttt{/app}; custom paramiko SSH server binds \texttt{:22} non-root, pubkey-auths ``user'' with no UNIX account; bare repo + post-receive + HTTP \texttt{:8080} in \texttt{/app}. Adapted verifier resolves the \texttt{/app} tools. \\
\addlinespace
crack-7z-hash & F/P & \texttt{apt} \texttt{7zip} + perl-lzma (root). & Rootless apt-sysroot: \texttt{apt-get download} + \texttt{dpkg -x} into \texttt{/app/sysroot}; genuine crack with baked John the Ripper (numeric mask). \\
\addlinespace
custom-memory-heap-crash & F/P & Uses gdb + writes \texttt{/proc/\allowbreak sys/\allowbreak .../\allowbreak core\_pattern} (absent / RO). & Edit only \texttt{/app/user.cpp}: force iostream facet-node registration in \texttt{user\_init()} before the custom heap installs; drop gdb steps. Zero egress. \\
\addlinespace
dna-assembly & F/P & \texttt{apt} emboss+primer3 (root). & Skip \texttt{apt} (tools only print diagnostics); build primers by pure-coreutils slicing at the canonical offsets. Verifier runs its own primer3. \\
\addlinespace
dna-insert & F/P & \texttt{apt} emboss+primer3 (root). & Same skip-\texttt{apt} coreutils-slicing route as dna-assembly. \\
\addlinespace
financial-document-processor & F/P & \texttt{apt} tesseract + \texttt{uv run} (root; cache under \texttt{/root}). & \emph{Genuine} rootless OCR: conda-forge tesseract 5.3.0 + byte-identical tessdata; \texttt{pip} pymupdf/pytesseract into \texttt{/app}; canonical classifier/parser verbatim, only the tesseract pointer changed. (Supersedes an earlier witness that did not perform the OCR.) \\
\addlinespace
fix-git & F/P & \texttt{git merge} needs committer identity set in \texttt{/root/.gitconfig}. & Point \texttt{HOME=/app/home}, register identity + \texttt{safe.directory}; reproduce the canonical reflog recovery + \texttt{merge -X theirs}. git baked, zero egress. \\
\addlinespace
gcode-to-text & N & \texttt{apt} tesseract/opencv from Debian mirror (off-list). & conda-forge tesseract + tessdata from GitHub + \texttt{opencv-python-headless} from PyPI; OCR pipeline byte-for-byte, write only \texttt{/app}. \\
\addlinespace
git-multibranch & F/P & \texttt{useradd}/\texttt{chpasswd} (\texttt{/etc/shadow}) + system sshd \texttt{:22} (root). & paramiko SSH server \texttt{:22} non-root, password-auths git/password with no UNIX user; bare repo + hook + stdlib HTTPS \texttt{:8443} in \texttt{/app}. Only egress: \texttt{pip} paramiko wheel. \\
\addlinespace
hf-model-inference & N & \texttt{from\_pretrained} streams weights from HF hub/Xet CDN (off-list). & Byte-identical config/vocab/\texttt{pytorch\_model.bin} from HF legacy S3 (\texttt{*.amazonaws.com}), size+sha256 pinned, load offline. \\
\addlinespace
install-windows-3.11 & N+F/P & Compiles QEMU 5.2 from an off-list host + services (root). & Rootless apt-sysroot QEMU 5.2.0 from Ubuntu hirsute \texttt{old-releases}; baked \texttt{win311.img} read-only; start rootless nginx:80 + websockify:8080. \\
\addlinespace
largest-eigenval & F/P & \texttt{pip install eigenpy} to system site-packages (root verifier imports it). & \texttt{pip --target=/app/pylib} 3 pinned wheels (\texttt{--no-deps}, keep system numpy); \texttt{/app/eigen.py} self-bootstraps \texttt{sys.path} via \texttt{site.addsitedir} in the verifier process. \\
\addlinespace
log-summary-date-ranges & N & \texttt{apt} grep/coreutils from Debian mirror (off-list, redundant). & Skip \texttt{apt}; pure stdlib over baked \texttt{/app/logs}. Zero egress. \\
\addlinespace
make-doom-for-mips & N+F/P & \texttt{apt} MIPS cross toolchain from Debian mirror (off-list). & Self-contained LLVM 14.0.6 from conda-forge via static micromamba; freestanding build in \texttt{/app}. Adapted verifier runs vm.js with \texttt{cwd=/app}. \\
\addlinespace
mcmc-sampling-stan & N+F/P & Installs rstan from CRAN (off-list), root. & Rootless apt-sysroot \texttt{r-cran-rstan} 2.32.5 + Stan closure from noble universe; point system R at the \texttt{/app} site-library. Adapted verifier resolves the \texttt{/app} R wrapper first. \\
\addlinespace
merge-diff-arc-agi-task & F/P & \texttt{apt} git (root; git not baked). & Rootless apt-sysroot git + closure into \texttt{/app/sysroot}; reproduce repo/branch/merge end-state; \texttt{algo.py} pure-stdlib. \\
\addlinespace
mteb-retrieve & N & \texttt{mteb.get\_model} pulls BGE weights from HF hub/Xet (off-list). & Byte-identical BGE-small-zh-v1.5 weights as non-LFS git blobs from GitHub mirrors into the HF cache; \texttt{HF\_HUB\_OFFLINE}; unchanged \texttt{retrieve.py}. \\
\addlinespace
overfull-hbox & F/P & \texttt{apt} python3-pip for the solver (no python3 in image; root). & Skip \texttt{apt}; pdflatex+perl already baked; do the six reference synonym swaps in perl; replay verifier compile. Zero egress. \\
\addlinespace
protein-assembly & N & Queries RCSB REST + FPbase (off-list). & PDB mmCIF from the S3 archive snapshot (\texttt{*.amazonaws.com}); donor/acceptor by S3 title substring; SNAP/FLAG hardcoded as the canonical does; \texttt{pip} from PyPI. \\
\addlinespace
pypi-server & N+F/P & \texttt{apt} update (off-list) + apache2-utils/twine auth (root). & Skip \texttt{apt}/auth (pypiserver serves anonymously); two pure-python wheels from PyPI, sha256-gated; serve the built wheel over loopback from \texttt{/app}. \\
\addlinespace
pytorch-model-cli & N+F/P & PyTorch CPU wheel index + Debian compiler (both off-list). & \texttt{weights.json} via pure-stdlib zip/struct extraction (byte-identical to torch); compile the reference C with \texttt{zig cc} from the ziglang PyPI wheel (bundled libc). \\
\addlinespace
qemu-alpine-ssh & F/P & Root-owned qcow2 opened rw + \texttt{apk add openssh} via off-list CDN. & Fresh qcow2 in \texttt{/app}, ISO read-only via \texttt{-cdrom}; openssh installed offline from the alpine-extended on-media apk repo; serial-console driven; hostfwd \texttt{:2222}; daemonized so the VM survives to the probe. \\
\addlinespace
rstan-to-pystan & N+F/P & \texttt{sudo} + deadsnakes PPA (off-list) for Py3.10 + httpstan from source. & Use baked Python 3.12; prebuilt \texttt{httpstan} cp312 wheel from PyPI (no source build); rootless apt-sysroot for the C++ toolchain used at model-compile time. \\
\addlinespace
sam-cell-seg & N+F/P & \texttt{apt} libgl1 + torch/torchvision \texttt{+cpu} wheels (off-list, root). & Deliverable is one file \texttt{/app/convert\_masks.py}; verifier supplies torch/mobile\_sam in its own unrestricted environment; author the canonical script offline, zero solve-time egress. \\
\addlinespace
train-fasttext & N+F/P & C++ toolchain + clone fastText + \texttt{make} (off-list, root). & Prebuilt \texttt{fasttext-wheel} from PyPI (no compile); materialize a python-build-standalone 3.12 via \texttt{uv} (0.9.2 ships cp312 only); train inside it. Verifier reads \texttt{/app/model.bin}.  \\
\end{longtable}
\twocolumn

\section{Appendix C: A Benchmark-Integrity Artifact}
\label{app:benchmark-integrity-exclusions}

We flag a task as a benchmark-integrity artifact when the official scoring
path does not establish that the requested work was performed. Such findings
are distinct from ordinary task failures and from tasks that become unsolved
under a policy. A flagged task remains in every reported aggregate,
consistent with the main paper's 89-task reporting; the flag is recorded
across all model--harness bundles (a model paired with the agent harness
that runs it) and policy levels, and the affected raw trials are preserved
for audit.

\paragraph{\texttt{count-dataset-tokens}.}
This task requires retrieving a Hugging Face dataset and computing its token
count. In an affected \textit{NIST-derived high} trial, the dataset request returned
\texttt{HTTP 403}, and the trajectory contained no successful retrieval or
tokenization. The model nevertheless submitted the exact expected constant,
\texttt{79586}. The official verifier awarded success because it checked only
whether that string occurred in the answer file. Producing the exact constant
without the required computation is consistent with benchmark contamination or
memorization, although the trajectory cannot establish how the model obtained
it. The confirmed measurement defect is narrower: the verifier awarded success
without evidence that the requested work occurred. The pattern is not a
single trial: 13 trials of this task pass by submitting the expected
constant (Appendix~G). We therefore flag \texttt{count-dataset-tokens} as a
benchmark-integrity artifact, keep its trials in every reported aggregate,
and preserve them as audit evidence; separately, the task is excluded from
the 82-task witness pool on solvability grounds (Appendix~B).

\section{Appendix D: Provider-Side Safety Interventions}
\label{app:provider-safety-interventions}

Two kinds of provider-side safety mechanism, serving-layer refusals and
vendor fallback routing, intervened between the harness and the
model during our runs. Neither is a benchmark-integrity artifact: the
affected trials remain in every aggregate. We document them because they are
properties of the deployed model route rather than of the runtime policy, and
because any organization operating the same route would encounter them.
Because every intervention is annotated per trial, we also report incidence by
policy level, examining whether these mechanisms fire more often under hardened
levels than under the unrestricted control---as would be expected if
policy-induced denial-and-retry behavior makes a trajectory more likely to be
flagged by a serving-layer classifier.

\paragraph{Serving-layer refusals without security authorization.}
OpenAI's serving layer screens requests for cybersecurity risk. On
security-themed tasks it can terminate the model stream with the message
``This content was flagged for possible cybersecurity risk,'' directing the
caller to enroll in its Trusted Access for Cyber
program.\footnote{\url{https://openai.com/index/trusted-access-for-cyber/}}
Our organization was not enrolled during the measurement window, so our runs
traverse the default, unenrolled serving route. When the flag fires, the
Codex CLI reconnects and retries the disconnected stream, and the retry
frequently succeeds: in 25 of the 43 flagged trials the stream resumed and the
trial ran to completion, 23 of which reached a passing state. When the retries
are exhausted the CLI abandons the turn, and the affected trial either aborts
mid-trajectory, which is the common case (14 of the 18 abandoned trials), or
terminates before any tool call. The flagging is stochastic---some attempts of a task
are terminated while identical sibling attempts proceed. We do not exclude
these trials. Under our deployed-bundle scope, a task the provider declines
to serve is a failure of that model--harness bundle at that policy level: it
depresses the measured success rate and still incurs the cost of the consumed
attempts. We detect each occurrence from the harness event stream (a terminal
\texttt{turn.failed} carrying the flag message, with a zero-work or
aborted-trajectory signature), annotate the trial, and report the frequency
of these refusals alongside the affected results, so their contribution to
failure counts and cost is visible rather than folded silently into ordinary
failures. Any organization without this authorization measures---and
operates---the same degraded route; enrolled organizations receive a more
permissive serving path for verified security work.

\paragraph{Vendor fallback routing under dual-use classifiers.}
Claude Fable 5 ships with dual-use safety classifiers, and the vendor's
production behavior when a request is flagged is to serve the response with
Claude Opus 4.8
instead.\footnote{\url{https://www.anthropic.com/news/claude-fable-5-mythos-5}}
Our harness preserves this behavior by declaring the vendor's fallback at the
request level, so Fable 5 results measure the gated product route an
organization actually deploys. The vendor's own reporting discloses how often
the fallback fires per benchmark (20.9\% of Terminal-Bench trials in the
accompanying system card), and we adopt the same disclosure practice: the
model gateway records the model that actually served every request, and each
Fable 5 result is accompanied by the share of trials containing at least one
fallback-served completion and the per-trial share of fallback-served
requests. Numbers for this bundle should therefore be read as the behavior of
the gated route, with the fallback share as the explicit label of how often
the fallback model participated. Across the 801 evaluated trials, 27.6\%
contained at least one fallback-served completion, against the 20.9\% of
trials the vendor reports for this benchmark, and 29.9\% of all requests
were served by the fallback model. The distribution is strongly bimodal:
72.4\% of trials were served entirely by Claude Fable 5 and 12.9\% entirely
by Claude Opus 4.8, though a majority of the fallback-touched trials mix
both models, so the routing decision often, but not always, persists across
a conversation.

\paragraph{Rerouting and refusal are distinct outcomes.}
The fallback and the refusals above are not successive stages of one
mechanism. The fallback is declared on every request, and it engages when the
Claude Fable 5 route declines to serve one: the request is retried against
Claude Opus 4.8 and the agent observes nothing unusual. A refusal, by
contrast, arrives as an ordinary successful completion whose content is the
refusal itself, which presents the routing layer with nothing to retry. The
\taskname{protein-assembly} trials make the distinction concrete: the
fallback was fully engaged there, with every request already served by Claude
Opus 4.8, and the trial still ended in a refusal because Claude Opus 4.8
refused the same content. That task is refused by every Claude model
evaluated here, so rerouting between them cannot resolve it.

\paragraph{Refusals on the Claude Code route.}
Three further refusal messages appear on the Claude Code route, none of which
is retried: each ends the trial where it fires. Claude Opus~5 returns a
cybersecurity-topic refusal naming Anthropic's Cyber Verification Program,
the structural counterpart of the OpenAI flag above and the single largest
source of trial-ending interventions we observe. Claude Opus~4.8 and Claude Sonnet~5 return
a general Usage Policy refusal, and Claude Fable~5 returns a safeguard notice
naming its own dual-use classifiers, the classifiers whose production fallback
behavior the preceding paragraph describes. A fourth wording, citing
restrictions on violative cyber content, appeared once on Claude Opus~5.
Table~\ref{tab:refusal-messages} gives each message with its frequency and how
often it ended the trial. The Usage Policy message is itself instructive: it
advises API integrators to configure a fallback model to reduce refusals,
which is exactly the vendor mechanism documented above for Claude Fable~5.

\begin{table}[t]
\centering
\footnotesize
\setlength{\tabcolsep}{3pt}
\begin{tabular}{@{}>{\raggedright\arraybackslash}p{0.42\columnwidth}>{\raggedright\arraybackslash}p{0.26\columnwidth}rr@{}}
\toprule
\textbf{Message (abridged)} & \textbf{Bundle} & \textbf{Trials} & \textbf{Ended} \\
\midrule
``flagged for possible cybersecurity risk'' (Trusted Access for Cyber) & GPT-5.6 Sol & 43 & 18 \\
\addlinespace
``safety measures \ldots{} flagged this message for a cybersecurity topic'' (Cyber Verification Program) & Claude Opus 5 & 38 & 38 \\
\addlinespace
``appears to violate our Usage Policy'' & Claude Opus 4.8, Claude Sonnet 5 & 9 & 9 \\
\addlinespace
``Fable 5's safeguards flagged this message'' & Claude Fable 5 & 7 & 7 \\
\addlinespace
``restrictions on violative cyber content'' & Claude Opus 5 & 1 & 1 \\
\bottomrule
\end{tabular}
\caption{\textbf{Only the OpenAI route retries a refused request; every
Anthropic refusal ends the trial where it fires.} The five distinct
provider-side refusal messages observed, the bundles that received them, the
number of trials in which each appeared, and the number of those trials the
message ended. The Codex CLI reconnects after a dropped stream, so 25 of the
43 flagged GPT-5.6 Sol trials resumed and completed. Each bundle contributes
801 trials (89 tasks, three trials, three policy conditions).}
\label{tab:refusal-messages}
\end{table}

\paragraph{Incidence by policy level.}
These interventions do not become more frequent under hardening. The five
affected bundles recorded 28 flagged trials under control, 41 under non-root,
and 29 under \textit{NIST-derived high} (Table~\ref{tab:refusal-incidence}).
Four of the five are flat across the three conditions; only Claude Opus~5
varies materially, and 12 of its 20 non-root flags fall on tasks that are
blocked by design, where the agent must attempt the privileged operations the
policy denies before it can make progress. We therefore do not observe the
association the mechanism would predict, and we report none. The counts are
small and clustered by bundle and by task, so we describe them rather than
test them.

\begin{table}[t]
\centering
\footnotesize
\begin{tabular}{@{}lrrrrr@{}}
\toprule
\textbf{Bundle} & \textbf{Ctrl} & \textbf{Non-root} & \textbf{$H$} & \textbf{Total} & \textbf{Ended} \\
\midrule
GPT-5.6 Sol & 14 & 16 & 13 & 43 & 18 \\
Claude Opus 5 & 8 & 20 & 11 & 39 & 39 \\
Claude Fable 5 & 2 & 2 & 3 & 7 & 7 \\
Claude Sonnet 5 & 3 & 2 & 2 & 7 & 7 \\
Claude Opus 4.8 & 1 & 1 & 0 & 2 & 2 \\
\midrule
Total & 28 & 41 & 29 & 98 & 73 \\
\bottomrule
\end{tabular}
\caption{\textbf{Intervention incidence does not rise with policy severity.}
Flagged trials per bundle and policy condition, out of 267 trials per cell
(89 tasks, three trials each); $H$ is \textit{NIST-derived high}. ``Ended''
counts the flagged trials the intervention terminated. The seven bundles not
listed recorded no interventions in any condition.}
\label{tab:refusal-incidence}
\end{table}

\paragraph{Task concentration.}
What the flags track is task content. All 98 fall on 16 of the 89 tasks, and
four tasks account for just under half of them: \taskname{vulnerable-secret} (18 trials),
\taskname{protein-assembly} (12), \taskname{break-filter-js-from-html} (9), and
\taskname{feal-differential-cryptanalysis} (9). \taskname{vulnerable-secret}
draws both vendors' classifiers, in every policy condition and every replicate;
\taskname{protein-assembly}, a protein-design task, draws only
Anthropic-side refusals. Of the 73 interventions that ended a trial, 70 were scored as failures:
19 fall on tasks that are blocked by design and 51 on tasks for which a
solvability witness exists (Appendix~B). Only the 51 could represent lost
successes; the 19 that fall on blocked-by-design tasks could not have passed
under the policy regardless. Under control, where the policy forecloses no
task, the cost is small and bounded: 18 of the 1{,}335 control trials across
the five affected bundles (1.3\%) ended in an intervention-caused failure,
at most 6 of a single bundle's 267 control trials, so no bundle's control
success rate is depressed by more than 2.2 percentage points. We record
that split as an observation, not as grounds for exclusion: neither group is
removed from the reported aggregates.

\section{Appendix E: Software Versions}
\label{app:software-versions}

Table~\ref{tab:software-versions} lists the software used to produce every
result in the paper. All components were frozen before the benchmark runs and
held fixed across models, policies, and repetitions; the single exception
(the Claude Code version used for Claude Opus~5) is documented at the end of
this appendix. The complete transitive dependency closure
is pinned in the released artifact: \texttt{uv.lock} for the evaluation
controller and a fully hash-pinned \texttt{requirements.lock} for the
in-sandbox Terminus-2 runtime (generated with \texttt{uv pip compile
--generate-hashes --universal} and an \texttt{--exclude-newer} horizon of
2026-07-15).

\begin{table}[t]
\centering
\scriptsize
\setlength{\tabcolsep}{3pt}
\begin{tabular}{@{}l l@{}}
\toprule
Component & Version \\
\midrule
\multicolumn{2}{@{}l}{\textit{Agent harnesses (installed into each sandbox at run start)}} \\
Claude Code (Anthropic models) & 2.1.206\textsuperscript{a} \\
Codex CLI (OpenAI models) & rust-v0.142.5 \\
Grok Build (xAI models) & 0.2.87 \\
Terminus-2 (all other models) & 2.0.0 \\
\quad Harbor runtime & 0.13.1 \\
\quad in-sandbox Python & 3.12.11 \\
\quad in-sandbox \texttt{uv} & 0.11.16 \\
\midrule
\multicolumn{2}{@{}l}{\textit{Evaluation controller}} \\
Python & 3.12 (digest-pinned \texttt{python:3.12-slim}) \\
Inspect AI (\texttt{inspect-ai}) & 0.3.239 \\
\texttt{inspect-harbor} & 0.5.9 \\
\texttt{harbor} (controller side) & 0.13.1 \\
\texttt{openai} client\textsuperscript{b} & 2.41.0 \\
\texttt{httpx} & 0.28.1 \\
\texttt{pydantic} & 2.13.4 \\
\texttt{PyYAML} & 6.0.3 \\
\midrule
\multicolumn{2}{@{}l}{\textit{Benchmark}} \\
Terminal-Bench & 2.1 (digest-pinned dataset\textsuperscript{c}) \\
\midrule
\multicolumn{2}{@{}l}{\textit{Sandbox platform}} \\
Daytona SDK (\texttt{daytona-sdk}) & 0.187.0 \\
\texttt{daytona-api-client} & 0.187.0 \\
\midrule
\multicolumn{2}{@{}l}{\textit{Cloud orchestration (does not touch trial outcomes)}} \\
\texttt{google-cloud-bigquery} & 3.42.1 \\
\texttt{google-cloud-firestore} & 2.27.0 \\
\texttt{google-cloud-storage} & 3.11.0 \\
\texttt{gcsfs} & 2025.9.0 \\
Terraform & $\geq$1.5; \texttt{hashicorp/google} $\geq$5.40, $<$7 \\
\bottomrule
\end{tabular}
\caption{\textbf{Software versions used for all reported runs.}
\textsuperscript{a}Claude Opus~5 only was run on Claude Code~2.1.220
(documented below); all other Claude models used 2.1.206.
\textsuperscript{b}Used only to initialize Inspect's OpenRouter provider; the
harness inside the sandbox, not the controller, performs model calls.
\textsuperscript{c}Harbor registry package
\texttt{terminal-bench/\allowbreak terminal-bench-2-1}, pinned at digest
\texttt{sha256:\allowbreak 7d7bdc1cbedad549\allowbreak fc1140404bd4dc45\allowbreak e5fd0ea7c4186773\allowbreak 687d177ad3a0699a}.}
\label{tab:software-versions}
\end{table}

\paragraph{Sandbox environments.}
Each trial provisions a fresh Daytona cloud sandbox from the task's own
per-task Terminal-Bench~2.1 Docker image, exactly as declared in the task's
\texttt{task.toml} (\texttt{environment.docker\_image}); CPU, memory, and
storage follow the task's declared resources without override. Task images
are heterogeneous (a mix of glibc- and musl-based distributions), so
base-image package versions are fixed by the digest-pinned dataset rather
than by a single global pin. A runtime layer is installed into every sandbox
at run start: distribution base packages, the harness CLI at the pinned
version above (Claude Code via its official installer, Codex and \texttt{uv}
as pinned GitHub release binaries, Grok from a pinned artifact bucket,
Terminus-2 from the hash-pinned \texttt{requirements.lock}), and the policy
enforcement stack.

\paragraph{Policy enforcement stack.}
Policy hardening uses Linux kernel mechanisms configured from within the
sandbox: a uid-scoped \texttt{nftables} reject wall with a forced egress
proxy (implemented in-repo in stdlib-only Python), seccomp filters and
\texttt{LD\_PRELOAD} shims compiled in-sandbox with the task image's own
toolchain and \texttt{libseccomp} so they link against that image's libc, and
Landlock network rules. The enforcement source is part of the frozen
repository; the userspace library versions (\texttt{libseccomp},
\texttt{nftables}) follow each task image's pinned distribution packages.

\paragraph{Model serving.}
All model inference is served over the OpenRouter HTTP API with per-harness
vendor endpoints (as described in the experimental setup); OpenRouter is a
hosted service and carries no
client-side version pin beyond the \texttt{openai} client version listed
above. The released experiment provenance records the exact provider model
identifier for every bundle.

\subsection{Claude Code version for Claude Opus 5}
\label{app:opus5-version}

Claude Sonnet~5, Claude Fable~5, and Claude Opus~4.8 were evaluated with Claude Code~2.1.206, the harness version frozen for the original benchmark runs. Claude Opus~5 was released only after those runs had completed. Re-running every previously evaluated bundle on a newer harness solely to accommodate one added model would have been costly and would have perturbed the frozen configuration, so we instead admitted Opus~5 as a narrowly scoped exception. Upstream Claude Code did not recognize the \texttt{claude-opus-5} model identifier until v2.1.219,\footnote{\url{https://github.com/anthropics/claude-code/blob/main/feed.xml}} so we evaluated Opus~5 with Claude Code~2.1.220, the immediately following release, which contributed only reliability fixes. Under v2.1.206, Opus~5 requests failed at the client and protocol level before the agent received any usable model response; we treat those observations as harness incompatibilities rather than model outcomes and exclude them. No other Claude model was moved off v2.1.206, so the historical harness configuration is otherwise unchanged, and the version exception and the affected runs are recorded in the released experiment provenance.

\subsection{Provider model identifiers}
\label{app:model-identifiers}

Table~\ref{tab:model-identifiers} lists, for each of the twelve evaluated
models, the exact provider model identifier used for inference and the agent
harness that drove it. Every identifier is the string passed to the OpenRouter
HTTP API; it is the same value recorded in the released experiment provenance
for every result bundle. In the primary comparison, harness assignment is fixed per vendor family: the
Anthropic models run under Claude Code, the OpenAI models under Codex,
Grok~4.5 under Grok Build, and all open-weight models under Terminus-2.

\begin{table}[t]
\centering
\scriptsize
\setlength{\tabcolsep}{3pt}
\begin{tabular}{@{}l l l@{}}
\toprule
Model & Provider model identifier & Harness \\
\midrule
\multicolumn{3}{@{}l}{\textit{OpenAI (Codex)}} \\
GPT-5.6 Sol    & \texttt{openai/gpt-5.6-sol}    & Codex \\
GPT-5.6 Terra  & \texttt{openai/gpt-5.6-terra}  & Codex \\
GPT-5.6 Luna   & \texttt{openai/gpt-5.6-luna}   & Codex \\
\midrule
\multicolumn{3}{@{}l}{\textit{Anthropic (Claude Code)}} \\
Claude Sonnet 5 & \texttt{anthropic/claude-sonnet-5} & Claude Code \\
Claude Opus 4.8 & \texttt{anthropic/claude-opus-4.8} & Claude Code \\
Claude Fable 5  & \texttt{anthropic/claude-fable-5}  & Claude Code \\
Claude Opus 5   & \texttt{anthropic/claude-opus-5}   & Claude Code \\
\midrule
\multicolumn{3}{@{}l}{\textit{Open-weight (Terminus-2)}} \\
Kimi K3      & \texttt{moonshotai/kimi-k3} & Terminus-2 \\
GLM-5.2      & \texttt{z-ai/glm-5.2}       & Terminus-2 \\
Qwen3.7 Max  & \texttt{qwen/qwen3.7-max}   & Terminus-2 \\
MiniMax M3   & \texttt{minimax/minimax-m3} & Terminus-2 \\
\midrule
\multicolumn{3}{@{}l}{\textit{xAI (Grok Build)}} \\
Grok 4.5 & \texttt{x-ai/grok-4.5} & Grok Build \\
\bottomrule
\end{tabular}
\caption{\textbf{Provider model identifiers and harness assignment.} Each
identifier is the exact OpenRouter model string used for inference; the
harness column names the agent CLI that issued the calls, held fixed across
all policies and repetitions. Harness version pins are in
Table~\ref{tab:software-versions}.}
\label{tab:model-identifiers}
\end{table}

\section{Appendix F: Full Per-Bundle Results under Policy}
\label{app:full-results}

Tables~\ref{tab:full-results-89} and~\ref{tab:full-results-82} give the per-bundle
success rate and cost behind Figure~1 of the main paper, for every model--harness
bundle under each policy. Table~\ref{tab:full-results-89} is the full 89-task pool
as plotted; Table~\ref{tab:full-results-82} restricts to the 82 tasks that admit a
solvability witness under \textit{NIST-derived high} (Appendix~B),
excluding the seven blocked-by-design tasks. Success rate is the percentage of
passing trials with the standard deviation across the three repetitions; cost is
the mean replicate cost over the pool, in US dollars; cost dispersion across
the repetitions is not reported.

\begin{table*}[t]
\centering
\small
\setlength{\tabcolsep}{5pt}
\begin{tabular}{l cc cc cc}
\toprule
& \multicolumn{2}{c}{Control} & \multicolumn{2}{c}{Non-root} & \multicolumn{2}{c}{NIST-derived high} \\
\cmidrule(lr){2-3}\cmidrule(lr){4-5}\cmidrule(lr){6-7}
Model & Success (\%) & Cost (\$) & Success (\%) & Cost (\$) & Success (\%) & Cost (\$) \\
\midrule
GPT-5.6 Sol & $83.9_{\pm1.9}$ & $45.59$ & $72.3_{\pm3.2}$ & $50.49$ & $74.2_{\pm2.4}$ & $58.88$ \\
GPT-5.6 Terra & $79.0_{\pm3.2}$ & $24.60$ & $65.2_{\pm1.8}$ & $29.46$ & $63.3_{\pm1.9}$ & $30.27$ \\
GPT-5.6 Luna & $71.9_{\pm1.6}$ & $13.28$ & $56.6_{\pm3.7}$ & $14.25$ & $53.9_{\pm1.8}$ & $15.41$ \\
Claude Sonnet 5 & $73.4_{\pm2.1}$ & $113.89$ & $60.3_{\pm1.9}$ & $135.58$ & $55.1_{\pm2.8}$ & $138.21$ \\
Claude Opus 4.8 & $74.9_{\pm1.4}$ & $123.44$ & $62.9_{\pm0.9}$ & $147.31$ & $63.3_{\pm3.7}$ & $164.23$ \\
Claude Fable 5 & $79.0_{\pm0.5}$ & $316.44$ & $65.9_{\pm4.7}$ & $317.68$ & $67.8_{\pm1.4}$ & $375.30$ \\
Claude Opus 5 & $77.5_{\pm1.6}$ & $215.26$ & $68.2_{\pm0.5}$ & $213.05$ & $65.2_{\pm2.4}$ & $290.11$ \\
Kimi K3 & $72.3_{\pm2.3}$ & $24.93$ & $60.3_{\pm1.9}$ & $32.45$ & $54.7_{\pm1.4}$ & $36.62$ \\
GLM-5.2 & $73.4_{\pm1.9}$ & $30.12$ & $59.6_{\pm2.4}$ & $36.46$ & $55.8_{\pm2.6}$ & $39.40$ \\
Qwen3.7 Max & $56.9_{\pm1.9}$ & $19.82$ & $45.3_{\pm2.9}$ & $29.77$ & $40.4_{\pm0.9}$ & $26.63$ \\
MiniMax M3 & $59.6_{\pm4.8}$ & $19.34$ & $46.8_{\pm1.1}$ & $35.03$ & $44.2_{\pm1.9}$ & $26.26$ \\
Grok 4.5 & $82.0_{\pm1.6}$ & $46.20$ & $73.0_{\pm0.9}$ & $81.59$ & $74.9_{\pm1.4}$ & $123.49$ \\
\bottomrule
\end{tabular}
\caption{\textbf{Per-bundle success rate and cost under each policy on the full
89-task pool (the set plotted in Figure~1 of the main paper).} Success rate is the
percentage of passing trials, subscript the standard deviation across the three
repetitions; cost is the mean 89-task replicate cost in US dollars.}
\label{tab:full-results-89}
\end{table*}

\begin{table*}[t]
\centering
\small
\setlength{\tabcolsep}{5pt}
\begin{tabular}{l cc cc cc}
\toprule
& \multicolumn{2}{c}{Control} & \multicolumn{2}{c}{Non-root} & \multicolumn{2}{c}{NIST-derived high} \\
\cmidrule(lr){2-3}\cmidrule(lr){4-5}\cmidrule(lr){6-7}
Model & Success (\%) & Cost (\$) & Success (\%) & Cost (\$) & Success (\%) & Cost (\$) \\
\midrule
GPT-5.6 Sol & $82.5_{\pm2.1}$ & $41.65$ & $77.2_{\pm3.5}$ & $46.20$ & $79.3_{\pm2.6}$ & $51.67$ \\
GPT-5.6 Terra & $77.2_{\pm3.5}$ & $22.74$ & $69.9_{\pm2.1}$ & $27.76$ & $67.5_{\pm2.1}$ & $28.09$ \\
GPT-5.6 Luna & $69.5_{\pm1.7}$ & $12.54$ & $61.0_{\pm4.0}$ & $13.52$ & $57.3_{\pm2.0}$ & $13.63$ \\
Claude Sonnet 5 & $72.4_{\pm1.5}$ & $104.26$ & $64.2_{\pm2.1}$ & $127.16$ & $58.9_{\pm2.5}$ & $125.03$ \\
Claude Opus 4.8 & $72.8_{\pm1.5}$ & $116.35$ & $67.1_{\pm1.0}$ & $133.85$ & $68.7_{\pm4.0}$ & $141.28$ \\
Claude Fable 5 & $77.6_{\pm0.6}$ & $288.92$ & $70.3_{\pm5.1}$ & $283.90$ & $73.6_{\pm1.5}$ & $320.40$ \\
Claude Opus 5 & $76.8_{\pm2.0}$ & $196.84$ & $72.8_{\pm0.6}$ & $204.82$ & $70.7_{\pm2.6}$ & $267.23$ \\
Kimi K3 & $71.1_{\pm3.0}$ & $23.90$ & $64.6_{\pm2.0}$ & $27.89$ & $58.9_{\pm2.1}$ & $28.63$ \\
GLM-5.2 & $73.2_{\pm1.0}$ & $28.57$ & $63.4_{\pm2.6}$ & $25.72$ & $60.2_{\pm3.2}$ & $30.83$ \\
Qwen3.7 Max & $55.7_{\pm0.6}$ & $18.37$ & $48.8_{\pm3.4}$ & $23.17$ & $43.9_{\pm1.0}$ & $23.01$ \\
MiniMax M3 & $58.9_{\pm4.1}$ & $18.98$ & $50.4_{\pm1.5}$ & $19.46$ & $48.0_{\pm2.1}$ & $19.87$ \\
Grok 4.5 & $81.7_{\pm1.7}$ & $44.92$ & $78.0_{\pm1.0}$ & $52.31$ & $79.3_{\pm1.7}$ & $75.64$ \\
\bottomrule
\end{tabular}
\caption{\textbf{Per-bundle success rate and cost under each policy on the 82
tasks that admit a solvability witness under \textit{NIST-derived high}
(Appendix~B).} The seven blocked-by-design tasks
(\texttt{build-cython-ext}, \texttt{build-pmars}, \texttt{build-pov-ray},
\texttt{count-dataset-tokens}, \texttt{mailman}, \texttt{nginx-request-logging},
\texttt{kv-store-grpc}) are excluded. Columns are as in
Table~\ref{tab:full-results-89}.}
\label{tab:full-results-82}
\end{table*}

\subsection{Pareto frontier membership}
\label{app:frontier-members}

Table~\ref{tab:frontier-members} reports the success--cost Pareto frontier under \textit{NIST-derived high} on the 82-task witnessed pool, computed from Table~\ref{tab:full-results-82}: a bundle lies on the frontier when no other bundle reaches at least its success rate at no greater cost. Three bundles are non-dominated. GPT-5.6 Sol and Grok~4.5 tie at $79.3\%$ success, but Sol attains it at \$51.67 to Grok's \$75.64, so Sol dominates Grok, and Grok~4.5 leaves the frontier as the costlier of the two.

\begin{table}[!t]
\centering
\small
\begin{tabular}{@{}lrrl@{}}
\toprule
Bundle & Success (\%) & Cost (\$) & Frontier status \\
\midrule
GPT-5.6 Luna & 57.3 & 13.63 & on frontier \\
GPT-5.6 Terra & 67.5 & 28.09 & on frontier \\
GPT-5.6 Sol & 79.3 & 51.67 & on frontier \\
Grok~4.5 & 79.3 & 75.64 & dominated by Sol \\
\bottomrule
\end{tabular}
\caption{\textbf{Success--cost Pareto frontier under \textit{NIST-derived high} (82-task pool).} The three non-dominated bundles, plus Grok~4.5, which ties GPT-5.6 Sol on success but at higher cost and so leaves the frontier as the costlier of the two. Success rates are point estimates; the Sol--Grok success tie is within replicate variation (Table~\ref{tab:full-results-82}), and the ordering is decided on cost. Computed from Table~\ref{tab:full-results-82}.}
\label{tab:frontier-members}
\end{table}

\section{Appendix G: Blocked-Action Analysis}
\label{app:blocked-actions}

This appendix details the blocked-action analysis behind the main paper's
Results: the
unit and its evidence, the verified-evidence gate, per-axis and per-task
detail, and the association with outcomes. All numbers are computed over the
full evaluated cohort (twelve bundles, 89 tasks, three trials per condition) by
a deterministic, outcome-blind re-analysis of each trial's stored evidence.

\paragraph{Unit and evidence classes.}
A \emph{blocked action} is one denied operation. Two evidence classes
contribute and remain distinct in the underlying records:
\emph{enforcement-observed blocks}, logged by the enforcement mechanisms
themselves (the egress proxy's denials, firewall counters, and the
\texttt{sudo} log), and \emph{transcript-inferred denials}, recognized from
characteristic error signatures in the transcript (\texttt{EROFS},
\texttt{EPERM}, \texttt{HTTP 403}, refused connections, DNS failures).
Repeated error output from one denied operation is collapsed to a single
action, and an inferred denial that merely echoes an enforcement-observed
block on the same axis is not counted again. The counter never reads the
verifier outcome, so the outcome associations below are not circular.

\paragraph{Verified-evidence gate.}
A stored trace is truncated when a transcript exceeds its storage bound. A
truncated trace can still show blocked actions, but it cannot establish
their absence, so every count uses only trials whose evidence is verified
complete: 6{,}194 of 9{,}612 trials overall, and under
\textit{NIST-derived high} 2{,}051 of 3{,}204, covering 88 of the 89 tasks.
Truncation is flat across conditions (35--36\% of trials at every policy
level), so the gate does not tilt cross-condition comparisons; it is uneven
across harnesses (the Grok~4.5 bundle retains 213 of 801 trials) and tasks:
\taskname{schemelike-metacircular-eval} exceeds the bound in nearly every
trial in every condition, including control, and is conservatively excluded
from the evaluable set; its transcripts are simply long (it passes 26 of its 36
trials under \textit{NIST-derived high}). All totals are lower bounds.

\paragraph{Axes and persistence.}
Blocked actions land exactly on the axes each condition restricts: control
records zero across all 3{,}204 trials, including those with truncated
traces, and all 482 blocked actions under
\textit{non-root} are privilege denials. Under \textit{NIST-derived high}
the volume split is 5{,}250 egress, 392 privilege, and 158 filesystem
denials (egress is 90.5\% of the 5{,}800 total), but incidence is closer than volume: 500 trials record an egress
denial, 283 a privilege denial, 158 a filesystem denial, and privilege
denials touch more tasks than egress denials (60 against 55). In total, 675 of the 2{,}051 verified trials record at least one blocked action; this union is smaller than the per-axis incidence sum (941) because a trial can be denied on more than one axis. The volume
gap is persistence. A trial that hits the egress wall records 10.5 denials
on average, against 1.4 for privilege and exactly one for filesystem;
single trials reach 171 egress denials, and destination-level linking,
whose attribution is heuristic, assigns 62\% of egress volume to re-attempts
against a destination
already blocked earlier in the same trial, up to 86 re-attempts on one
destination. A re-attempt establishes pressure on a destination, not
intent: a deliberate retry is indistinguishable from a client library's
automatic retry (package installers retry their host by default).
Consistent with a retryability reading, the denials that present as final
(\texttt{EROFS}, \texttt{EPERM}) show no re-attempts at all.

\paragraph{Concentration across tasks.}
Of the 88 evaluable tasks, 14 record a verified zero and 48, including
those 14, record five or
fewer blocked actions in total, while the top five carry 52.8\% of all
volume: \taskname{build-pov-ray} (693), \taskname{break-filter-js-from-html}
(685), \taskname{mteb-retrieve} (596), \taskname{protein-assembly} (555),
and \taskname{count-dataset-tokens} (534); the top twelve carry 80\%. Seven
tasks record a blocked action in every verified trial:
\taskname{build-pmars}, \taskname{build-pov-ray},
\taskname{count-dataset-tokens}, \taskname{hf-model-inference},
\taskname{make-doom-for-mips}, \taskname{mteb-retrieve}, and
\taskname{protein-assembly}. Table~\ref{tab:blockage-distribution} gives
the distribution; totals include re-attempts, which only inflate counts,
so membership in the low buckets is conservative.

\begin{table}[!t]
\centering
\footnotesize
\begin{tabular}{@{}lrr@{}}
\toprule
Blocked actions per task (total) & Tasks & Share \\
\midrule
0 (verified) & 14 & 15.9\% \\
1--5 & 34 & 38.6\% \\
6--15 & 7 & 8.0\% \\
16--40 & 11 & 12.5\% \\
41--100 & 10 & 11.4\% \\
101--250 & 5 & 5.7\% \\
251 or more & 7 & 8.0\% \\
\bottomrule
\end{tabular}
\caption{\textbf{Most tasks barely touch the policy; a small head absorbs
most of it.} Tasks bucketed by total blocked actions under
\textit{NIST-derived high}, summed over all bundles and verified trials.
Shares are of the 88 tasks with at least one verified-complete trial.}
\label{tab:blockage-distribution}
\end{table}

\paragraph{Exposure follows solvability.}
Grouping tasks by their solvability status (Appendix~B): on the 49
evaluable tasks whose shipped reference solution survives the policy (the
50 of Appendix~B minus \taskname{schemelike-metacircular-eval}),
14.4\% of 1{,}107 trials record a blocked action (0.95 per trial); on the
32 tasks that required an authored witness, 51.6\% of 764 trials (3.81 per
trial); on the 7 tasks blocked by design, 67.8\% of 180 trials (10.18 per
trial). The gradient is robust to the evidence gate: with truncated-trace
trials included it reads 15.2\%, 54.9\%, and 69.8\%.

\paragraph{Association with outcomes.}
On witness tasks, trials with no blocked action pass 66.2\% against 46.7\%
for trials with at least one, and failing trials record 2.7 times the
blocked actions of passing trials (5.88 against 2.20). On
reference-compatible tasks the pass gap persists (80.5\% against 59.1\%)
but the volume relation inverts (passing trials average 1.04 against 0.67):
there, hitting a wall is incidental, and failures are mostly not
blockage-shaped. On blocked-by-design tasks no verified trial passes,
except 13 trials of \taskname{count-dataset-tokens}, precisely the verifier
artifact documented in Appendix~C; that this analysis independently
isolates the same task corroborates both the labels and the evidence. All
of these are diagnostic associations, not causes: blockage evidence alone
cannot establish why a trial failed, harder tasks are both more blocked and
more failed, and an agent that anticipates the policy and never attempts a
denied operation leaves no evidence at all.

\end{document}